\documentclass[12pt,a4paper]{iopart}
\usepackage[english]{babel}

\usepackage{iopams}
\usepackage{verbatim}
\usepackage{color}

\usepackage{graphicx}

\usepackage{color}

\usepackage[colorlinks,citecolor=blue,linkcolor=blue,urlcolor=blue]{hyperref}

\def\iotabar{\lower3pt\hbox{$\mathchar'26$}\mkern-7mu\iota}
\newcommand {\aplt}{\ {\raise-.5ex\hbox{$\buildrel<\over\sim$}}\ }
\newcommand{\dd}{\mbox{d}}

\newcommand{\spe}{{\sigma}}

\newcommand{\eq}[1]{(\ref{#1})}
\newcommand{\bun}{\hat{\mathbf{b}}}

\newcommand{\bv}{\mathbf{v}}

\newcommand{\bR}{\mathbf{R}}

\newcommand{\bJ}{\mathbf{J}}

\newcommand{\bB}{\mathbf{B}}

\newcommand{\dotcross}{ \raise 0.65ex\hbox{${\scriptstyle {{_{\displaystyle \cdot}}\atop\times}}$} }
\newcommand{\crossdot}{ \raise 0.5ex\hbox{${\scriptstyle {{_\times}\atop{\displaystyle \cdot}}}$} }

\newcommand{\kappabf}{\mbox{\boldmath$\kappa$}}

\newcommand{\sumsig}{ \raise -1.3ex\hbox{${{\displaystyle \sum}\atop{\scriptstyle \sigma}}$} }
\newcounter{appnumb}

\begin{document}

\title[Optimizing stellarators for large flows]
{Optimizing stellarators for large flows}
\author{Iv\'an Calvo$^{1}$}
\vspace{-0.2cm}
\eads{\mailto{ivan.calvo@ciemat.es}}
\vspace{-0.5cm}
\author{Felix I Parra$^{2,3}$}
\vspace{-0.2cm}
\eads{\mailto{felix.parradiaz@physics.ox.ac.uk}}
\vspace{-0.5cm}
\author{J Arturo Alonso$^{1}$}
\vspace{-0.2cm}
\eads{\mailto{arturo.alonso@externos.ciemat.es}}
\vspace{-0.5cm}
\author{Jos\'e Luis Velasco$^{1}$}
\vspace{-0.2cm}
\eads{\mailto{joseluis.velasco@ciemat.es}}

\vspace{1cm}

\address{$^1$Laboratorio Nacional de Fusi\'on, CIEMAT, 28040 Madrid, Spain}
\address{$^2$Rudolf Peierls Centre for Theoretical Physics, University of Oxford, Oxford, OX1 3NP, UK}
\address{$^3$Culham Centre for Fusion Energy, Abingdon, OX14 3DB, UK}

\pacs{52.30.Gz, 52.35.Ra, 52.55.Hc}

\vskip 1cm

{\large
\begin{center}
\today
\end{center}
}

\vspace{1cm}

\begin{abstract}
  Plasma flow is damped in stellarators because they are not
  intrinsically ambipolar, unlike tokamaks, in which the flux-surface
  averaged radial electric current vanishes for any value of the
  radial electric field. Only quasisymmetric stellarators are
  intrinsically ambipolar, but exact quasisymmetry is impossible to
  achieve in non-axisymmetric toroidal configurations. By calculating
  the violation of intrinsic ambipolarity due to deviations from
  quasisymmetry, one can derive criteria to assess when a stellarator
  can be considered quasisymmetric in practice, i.e. when the flow
  damping is weak enough. Let us denote by $\alpha$ a small parameter
  that controls the size of a perturbation to an exactly
  quasisymmetric magnetic field. Recently, it has been shown that if
  the gradient of the perturbation is sufficiently small, the
  flux-surface averaged radial electric current scales as $\alpha^2$
  for any value of the collisionality. It was also argued that when
  the gradient of the perturbation is large, the quadratic scaling is
  replaced by a more unfavorable one. In this paper, perturbations
  with large gradients are rigorously treated. In particular, it is
  proven that for low collisionality a perturbation with large
  gradient yields, at best, an $O(|\alpha|)$ deviation from
  quasisymmetry. Heuristic estimations in the literature incorrectly
  predicted an $O(|\alpha|^{3/2})$ deviation.
\end{abstract}

\maketitle

\section{Introduction}
\label{sec:Introduction}

Quasisymmetry~\cite{Boozer83, Nuehrenberg88} is an attractive property
in stellarator design. It defines stellarator magnetic field
configurations that make these devices behave like tokamaks to lowest
order. In particular, the plasma flow is not damped in quasisymmetric
stellarators. Therefore, a perfectly quasisymmetric stellarator would
bring together the advantages of both concepts~\cite{Helander2012}. On
the one hand, the good confinement properties and rotation
capabilities of the tokamak. On the other hand, the steady state
character and absence of disruptions of the
stellarator. Some of the benefits of rotation, such
  as the stabilization of macroscopic magnetohydrodynamic
  modes~\cite{Vries96}, might not be as relevant in stellarators as
  they are in tokamaks because those modes are less worrisome in the
  former. But differential rotation, i.e. flow shear, can also reduce
  turbulent transport~\cite{Mantica09} and this is especially
  important at the edge of stellarators, where transport is dominated
  by turbulence~\cite{Dinklage13} and other stabilizing mechanisms
  like sheared rotational transform are typically not present. Looking
  for magnetic configurations that admit large flows gives a sensible
  path to eventually achieving large flow shear. Hence, requiring that
  the stellarator be quasisymmetric at least in a neighborhood of the
  edge seems a justified design goal.

Actually, exact quasisymmetry can only be achieved on
  a flux surface. Garren and Boozer proved in reference
  \cite{Garren1991} that the rest of the plasma volume will
  necessarily break the quasisymmetry condition. This does not
  invalidate the atractiveness of quasisymmetry as a stellarator
  optimization concept, however, because it is possible to reach
configurations reasonably close to
quasisymmetric~\cite{Anderson1995}. In other words, when quasisymmetry
is involved, one is necessarily dealing with a magnetic field that at
most has the form $\bB = \bB_0 + \alpha \bB_1$, where $\bB_0$ is
quasisymmetric and $\alpha \bB_1$ is a small deviation from
quasisymmetry. It seems important to understand how the desirable
properties of quasisymmetric stellarators are affected by the
perturbation $\alpha \bB_1$. The appropriate analysis of this problem
leads to the derivation of criteria to assess when a stellarator can
be called quasisymmetric in practice. In reference
\cite{CalvoParraVelascoAlonso13} we gave one such criterion for a wide
class of perturbations $\alpha \bB_1$.

A magnetic field is quasisymmetric if and only if the flux-surface
averaged lowest-order radial electric current identically vanishes for
any value of the radial electric field, and for any density and
temperature profiles~\cite{CalvoParraVelascoAlonso13,Helander08}. This
feature is called intrinsic ambipolarity and can be employed as the
defining property of quasisymmetry. In reference
\cite{CalvoParraVelascoAlonso13} we studied how the flux-surface
averaged radial electric current goes to zero as a function of
$\alpha$. We showed that, whenever the gradient of $B_0$ is much
larger than the gradient of $\alpha B_1$, the flux-surface averaged
radial electric current scales with $\alpha^2$, i.e.
\begin{equation}\label{eq:quadraticScaling}
  \left\langle
    \bJ\cdot\nabla\psi
  \right\rangle_\psi\sim\alpha^2k,
\end{equation}
where $\bJ$ is the electric current density, $\psi$ is the
flux-surface label coordinate, $\langle\cdot\rangle_\psi$ denotes the
flux-surface average operation defined in Section
\ref{sec:DeviationsFromQS} and the form of the factor $k$ depends on
the collisionality regime. For example,
\begin{equation}
k\sim \epsilon_i^2 e n_{i}v_{ti}|\nabla\psi|
\end{equation}
when $\nu_i L_0 / v_{ti}\sim 1$, whereas
\begin{equation}\label{eq:klowcollisionality}
k\sim \frac{\epsilon_i^2 v_{ti}}{\nu_i L_0}
 e n_i v_{ti}|\nabla\psi|
\end{equation}
when $\nu_i L_0 / v_{ti} \ll 1$. Here, $\epsilon_i := \rho_i/L_0$ is
the ion Larmor radius $\rho_i$ over the typical variation length of
$B_0$, $L_0 :=|\nabla \ln B_0|^{-1}$, $e$ is the proton charge,
$n_{i}$ is the equilibrium ion density, $v_{ti}$ is the ion thermal
speed and $\nu_i$ is the ion-ion collision frequency. It is worth
being more precise about the conditions under which this quadratic
scaling in $\alpha$ is obtained. Assume that $\{\psi,\Theta,\zeta\}$
are Boozer coordinates~\cite{Boozer81}, which exist as long as
$(\nabla\times\bB)\cdot\nabla\psi = 0$.  It is
known~\cite{Helander08} that in these coordinates $B_0\equiv
B_0(\psi,M\Theta-N\zeta)$ depends only on a single helicity
$M\Theta-N\zeta$ for some pair $(M,N)$. Without loss of generality we
can take $B_1(\psi,\Theta,\zeta)$ such that it has vanishing
flux-surface average and such that it does not contain the helicity
$(M,N)$. Then, the scaling \eq{eq:quadraticScaling} holds if
\begin{eqnarray}\label{eq:conditionsQuadraticScaling}
  \frac{|\alpha\partial_\Theta B_1|}{|\partial_\Theta B_0|}
\sim \alpha
,\nonumber\\[5pt]
\frac{|\alpha\partial_\zeta B_1|}
{|\partial_\zeta  B_0|}\sim \alpha
.
\end{eqnarray}

References \cite{Helander08} and \cite{Simakov11} proved that
flows are undamped in a stellarator only if it is
  intrinsically ambipolar, which is equivalent to being
  quasisymmetric. Then, it is easy to derive a criterion for rotation
  from \eq{eq:quadraticScaling}. This has been done in detail in
  \cite{CalvoParraVelascoAlonso13}. The result is that when
  \eq{eq:conditionsQuadraticScaling} holds, rotation in the symmetry
  direction can be close to sonic as long as
\begin{equation}\label{eq:rotationcriterion}
|\alpha| <
\epsilon_i^{1/2}.
\end{equation}

In order to avoid confusion, we need to discuss the result in
\cite{Sugama11}. In that reference, it has been proven that strictly
sonic equilibrium flows cannot take place in a stellarator, even if it
is quasisymmetric. However, the obstructions are absent if $\epsilon_i
v_{ti}\ll M v_{ti}\ll v_{ti}$, where $M = V_i/v_{ti}$ is the Mach
number and $V_i$ is the equilibrium plasma flow velocity. This is the
ordering in which our work should be understood.

It has also been argued in reference \cite{CalvoParraVelascoAlonso13}
that if \eq{eq:conditionsQuadraticScaling} is not satisfied, then the
scaling is expected to be more unfavorable than
\eq{eq:quadraticScaling}. Actually, an arbitrary perturbation that
satisfies $|\alpha| B_1 \ll B_0$ and, in addition,
  $|\alpha\partial_\Theta B_1|\sim|\partial_\Theta B_0|$ and
  $|\alpha\partial_\zeta B_1|\sim |\partial_\zeta B_0|$ (compare with
  \eq{eq:conditionsQuadraticScaling}) gives a flux-surface averaged
radial electric current $O(\alpha^0)$, and therefore the perturbed
configuration is not close to quasisymmetry anymore. The reason is
that when $|\alpha\partial_\Theta
  B_1|\sim|\partial_\Theta B_0|$ or $|\alpha\partial_\zeta B_1|\sim
  |\partial_\zeta B_0|$ one cannot guarantee that the radial
component of the magnetic drift,
$v_{\psi,\spe}:=\bv_{M,\spe}\cdot\nabla\psi$, can be expanded as
$v_{\psi,\spe} = v_{\psi,\spe}^{(0)} + \alpha v_{\psi,\spe}^{(1)} +
\dots$, where $v_{\psi,\spe} - v_{\psi,\spe}^{(0)} = O(\alpha)$ and
$v_{\psi,\spe}^{(0)}$ corresponds to the quasisymmetric magnetic
field. Since $v_{\psi,\spe}$ enters the drift kinetic equation as a
source term, $v_{\psi,\spe} - v_{\psi,\spe}^{(0)}$ will in general
yield a perturbation $O(\alpha^0)$ of the distribution function and
thus $\left\langle \bJ\cdot\nabla\psi \right\rangle_\psi =
O(\alpha^0)$.

The above results tell us that, when designing a stellarator that
intends to be quasisymmetric, it would be desirable to satisfy
\eq{eq:quadraticScaling}. If this is not possible due to other design
constraints, the breakdown of the $\alpha^2$ scaling does not
necessarily imply an $\alpha^0$ scaling. An intermediate result
between the $\alpha^2$ and $\alpha^0$ scalings is obtained when
$|\alpha\bun\cdot\nabla B_1| \sim |\bun\cdot\nabla B_0|$ but
\begin{equation}\label{eq:MagDriftCanBeExpanded}
  v_{\psi,\spe} - v_{\psi,\spe}^{(0)} = O(\alpha).
\end{equation}
Since $ v_{\psi,\spe} - v_{\psi,\spe}^{(0)} \propto (\bun\times\nabla
B_1)\cdot\nabla\psi$, condition \eq{eq:MagDriftCanBeExpanded} is
achieved for perturbations that satisfy $(\bun\times\nabla
B_1)\cdot\nabla\psi = O(\alpha)$, i.e. the component
  of $\nabla B_1$ along the flux surface is mostly parallel to the
  magnetic field lines. From now on and throughout the paper, we
assume that the stellarator has been designed so that
\eq{eq:MagDriftCanBeExpanded} is satisfied. Our objective is to find
out what scaling replaces \eq{eq:quadraticScaling}. Advancing the
final result, we will learn that for low collisionality the quadratic
scaling is replaced by
\begin{equation}\label{eq:linearScaling}
  \left\langle
    \bJ\cdot\nabla\psi
  \right\rangle_\psi\sim
\frac{|\alpha| \epsilon_i^2 v_{ti}}{L_0\nu_i} e n_i v_{ti} |\nabla\psi|.
\end{equation}

Equation \eq{eq:linearScaling} might seem surprising if one notes that
$|\alpha\bun_0\cdot\nabla B_1| \sim |\bun_0\cdot\nabla B_0|$ implies
that secondary wells can be created. It has typically been argued in
the literature~\cite{Ho1987} (see also \cite{Beidler2011} and
references therein) that these wells give a scaling of the radial
fluxes with $|\alpha|^{3/2}$ and that they dominate transport. We will
show that this is incorrect: particles trapped in secondary wells and
in large wells are both associated to a $|\alpha|$ scaling.

The rest of the paper is organized as follows. Section
\ref{sec:DeviationsFromQS} is a brief reminder of the derivation of
the $\alpha^2$ scaling given in \cite{CalvoParraVelascoAlonso13}. In
Section \ref{sec:breakdownOfExpansion} we explain in more detail why
this scaling can be broken when $|\alpha\bun_0\cdot\nabla B_1| \sim
|\bun_0\cdot\nabla B_0|$. We also show that in order to find the
scaling that replaces \eq{eq:quadraticScaling}, passing particles are
irrelevant and we can focus on trapped particles, distinguishing
between the ones trapped in large wells and those trapped in small
secondary wells. As a preliminary step, we work out the scaling of the
orbit-averaged radial magnetic drift for both types of trajectories in
Section \ref{sec:AverageMagDrift}. In Section \ref{sec:Matching} we
obtain the scaling of the distribution function and finally prove
\eq{eq:linearScaling}. We will also comment on the modification of the
rotation criterion \eq{eq:rotationcriterion}. The conclusions are
presented in Section \ref{eq:Conclusions}.

\section{Small helicity perturbations and $\alpha^2$ scaling}
\label{sec:DeviationsFromQS}

In this section we present the equations involved in our problem and
recall the results of Section 7 in reference
\cite{CalvoParraVelascoAlonso13}.

We employ phase-space coordinates $\{\bR,u,\mu\}$, with $\bR$ the
guiding-center position, $u$ the parallel velocity and $\mu$ the
magnetic moment. The drift-kinetic equations rely on the smallness of
the normalized ion Larmor radius, $\epsilon_i \ll 1$. The distribution
function is expanded as $F_{\spe} = F_{\spe 0} + F_{\spe
  1}+O(\epsilon_\spe^2 F_{\spe 0})$, with $F_{\spe 1}/F_{\spe 0} =
O(\epsilon_\spe)$. Here, $\epsilon_\spe = \rho_\spe/L_0$ is the ratio
of the Larmor radius of species $\spe$ over the typical variation
length of $B_0$. The condition $\epsilon_\spe\ll 1$ means that species
$\spe$ is strongly magnetized. The electrostatic potential is
expressed as $\varphi = \varphi_0 + \varphi_1 + O(\epsilon_i^2
T_i/e)$, where $\varphi_1/\varphi_0 = O(\epsilon_i)$. We adopt a
maximal expansion in which $\nu_{*\spe}\sim 1$, where $\nu_{*\spe} =
\nu_\spe L_0/v_{t\spe}$ is the collisionality of species $\spe$,
$v_{t\spe}$ is the thermal speed, and $\nu_\spe =
\sum_{\spe'}\nu_{\spe\spe'}$, $\nu_{\spe\spe'}$ is the frequency of
collisions between species $\spe$ and $\spe'$. To lowest order in
$\epsilon_i$ we deduce that $\varphi_0$ only depends on $\psi$ and
that $F_{\spe 0}$ is Maxwellian,
\begin{eqnarray}
&& \fl F_{\spe 0}(\bR,u,\mu)=
n_{\spe}
\left(\frac{m_\spe}{2\pi T_{\spe}}\right)^{3/2}
\exp\left(-\frac{m_\spe(u^2/2 + \mu B)}{T_{\spe}}\right),
\end{eqnarray}
where $m_\spe$ is the mass of species $\spe$, the density $n_\spe$ and
temperature $T_\spe$ depend only on $\psi$, and $T_\spe = T_{\spe'}$
for every pair $\spe$, $\spe'$ (the ion and electron temperatures can
be decoupled if a mass ratio expansion $\sqrt{m_e/m_i}\ll 1$ is
performed). The densities satisfy the lowest-order quasineutrality
equation, $\sum_\spe Z_\spe e n_\spe = 0$.

Define the non-adiabatic piece of the distribution function by
$G_{\spe 1}:= F_{\spe 1} + (Z_\spe e\varphi_1/T_\spe) F_{\spe 0}$. It
satisfies the {\it drift kinetic equation}
\begin{eqnarray}\label{eq:FPG1dimensionful}
\fl
\left(u\bun\cdot\nabla - \mu\bun\cdot\nabla
B\partial_u
\right)
G_{\spe 1}
\nonumber\\[5pt]
\fl\hspace{1cm}
+\Upsilon_\spe
\bv_{M,\spe}\cdot\nabla\psi
 F_{\spe 0}
= C^\ell_\spe[G_1].
\end{eqnarray}
Here, $C^\ell_\spe[G_1]$ is the linearized Fokker-Planck collision
operator,
\begin{eqnarray}\label{eq:vmDimensionful}
  \bv_{M,\spe} = \frac{1}{\Omega_\spe}\bun\times\big( 
u^2 \kappabf + \mu\nabla B\big)
\end{eqnarray}
is the magnetic-drift velocity, $\Omega_\spe = Z_\spe e B/(m_\spe c)$
is the gyrofrequency of species $\spe$, $c$ is the speed of light, and
\begin{eqnarray}
\fl
\Upsilon_\spe := \frac{Z_\spe e }{T_\spe}\partial_\psi\varphi_0 +
\frac{1}{n_\spe}\partial_\psi n_\spe
\nonumber\\[5pt]
\fl\hspace{1cm}
+
\left(\frac{m_\spe(u^2/2 +\mu B)}{T_\spe}-\frac{3}{2}\right)
\frac{1}{T_\spe}\partial_\psi T_\spe.
\end{eqnarray}

The neoclassical expression for the flux-surface average of the radial
electric current reads
\begin{eqnarray}\label{eq:FSAofradialJdimensionful}
\fl
\left\langle
\bJ\cdot\nabla\psi
\right\rangle_\psi
=
2\pi \left\langle
\sum_{\spe}
Z_\spe e
\int B
\bv_{M,\spe}\cdot\nabla\psi
\, G_{\spe 1}
\dd u \dd\mu\right\rangle_\psi,
\end{eqnarray}
where $Z_\spe e$ is the charge of species $\spe$. The flux-surface
average of a function $f(\psi,\Theta,\zeta)$ is
\begin{eqnarray}
\langle
f
\rangle_\psi
=
\frac{1}{V'}\int_0^{2\pi}\int_0^{2\pi} \sqrt{g} f\,\dd\Theta\dd\zeta.
\end{eqnarray}
For the moment, $\{\psi,\Theta,\zeta\}$ are arbitrary flux
coordinates, $\sqrt{g}$ is the square root of the metric determinant,
$V(\psi)$ is the plasma volume enclosed by the surface labeled by
$\psi$ and its derivative is given by
\begin{eqnarray}
V'(\psi)
=
\int_0^{2\pi}\int_0^{2\pi} \sqrt{g}\,\dd\Theta\dd\zeta.
\end{eqnarray}

The {\it
  ambipolarity condition},
\begin{eqnarray}\label{eq:AmbipolarityCondition}
\fl
2\pi \left\langle
\sum_{\spe}
Z_\spe e
\int B
\bv_{M,\spe}\cdot\nabla\psi
\, G_{\spe 1}
\dd u \dd\mu\right\rangle_\psi = 0,
\end{eqnarray}
imposes $\left\langle \bJ\cdot\nabla\psi \right\rangle_\psi$ to
vanish to lowest order in $\epsilon_i$. Equations
\eq{eq:FPG1dimensionful} and \eq{eq:AmbipolarityCondition} are the
relevant ones in stellarator neoclassical
calculations\footnote{To be precise, the neoclassical
    description of some low collisionality stellarator regimes, such
    as the $\sqrt{\nu}$ and $\nu$ regimes, requires additional terms
    in \eq{eq:FPG1dimensionful}. The reason is that at low
    collisionality $G_{\spe1}$ scales with $\nu_{*\spe}^{-1}$ and
    terms that are nominally of higher order in the $\epsilon_\spe$
    expansion of the drift-kinetic equation may actually be
    non-negligible.}.

We write our magnetic field as $\bB = \bB_{0} + \alpha \bB_1$,
where $\bB_0$ is quasisymmetric and $\alpha \bB_1$ is a small
perturbation. We assume that $\{\psi,\Theta,\zeta\}$ are Boozer
coordinates and take $B_1$ as explained below equation
\eq{eq:klowcollisionality}. We want to show that if
\eq{eq:conditionsQuadraticScaling} is satisfied, then $\left\langle
  \bJ\cdot\nabla\psi \right\rangle_\psi = O(\alpha^2)$. We recall
that in Boozer coordinates $\bB$ can be written as
\begin{eqnarray}
\fl
\bB = -\tilde\eta\nabla\psi + \frac{I(\psi)}{2\pi}\nabla\Theta +
\frac{J(\psi)}{2\pi}\nabla\zeta
\end{eqnarray}
and as
\begin{eqnarray}
\fl
\bB = \frac{\Psi'_p(\psi)}{2\pi}\nabla\zeta\times\nabla\psi +
\frac{\Psi'_t(\psi)}{2\pi}\nabla\psi\times\nabla\Theta.
\end{eqnarray}
The prime denotes differentiation with respect to $\psi$,
$\Psi_t$ is the toroidal flux, $\Psi_p$ the poloidal flux, and
$\tilde\eta(\psi,\Theta,\zeta)$ is a singly-valued function. An
important property of Boozer coordinates is that $\sqrt{g}$ can be
expressed in terms of the magnitude of the magnetic field,
\begin{equation}\label{eq:sqrtgBoozer}
\sqrt{g} = \frac{V' \langle B^2 \rangle_\psi}{4\pi^2 B^2}.
\end{equation}
The following related identity will be useful later on. Namely,
\begin{equation}\label{eq:bdotnablaTheta}
\bun\cdot\nabla\Theta
=
\frac{2\pi\Psi'_p B}{V'\langle B^2\rangle_\psi}.
\end{equation}

The derivative along the magnetic field reads
\begin{eqnarray}\label{eq:gradparBoozer}
\bun\cdot\nabla = \frac{2\pi\Psi'_t B}{\langle B^2 \rangle_\psi V'}
(\iotabar\partial_\Theta + \partial_\zeta),
\end{eqnarray}
where $\iotabar(\psi) = \Psi'_p(\psi)/\Psi'_t(\psi)$ is the rotational
transform. Finally, the radial component of the magnetic drift is
given by
\begin{equation}\label{eq:vpsiBoozer2}
\fl
v_{\psi, \spe}
:= \bv_{M,\spe}\cdot\nabla\psi
=
\frac{2\pi m_\spe c (u^2 + \mu B)}
{  Z_\spe e V'\langle B^2 \rangle_\psi  B}
\big(
I
\partial_\zeta B
-J \partial_\Theta B
\big),
\end{equation}
where $(\nabla\times\bB)\cdot\nabla\psi \equiv
0$ has been used. Therefore,
\begin{eqnarray}\label{eq:FSAofradialJdimensionfulBoozer}
\fl
\left\langle
\bJ\cdot\nabla\psi
\right\rangle_\psi
=
\nonumber\\[5pt]
\fl\hspace{0.5cm}
\sum_\spe \frac{m_\spe c}{V'}
\int_0^{2\pi}\int_0^{2\pi}
\dd\Theta\dd\zeta
\int
\frac{u^2 + \mu B}{B^2}
(I\partial_\zeta B - J \partial_\Theta B)
\, G_{\spe 1}
\dd u \dd\mu.
\end{eqnarray}
Observe equations \eq{eq:FPG1dimensionful}, \eq{eq:gradparBoozer},
\eq{eq:vpsiBoozer2}, \eq{eq:FSAofradialJdimensionfulBoozer}, and
recall that the kernel of the collision operator in drift-kinetic
coordinates depends on the magnetic field exclusively through $B$ (see
Appendix G of reference \cite{CalvoParraVelascoAlonso13}). Then, the
magnetic geometry information enters the drift kinetic equation and
the ambipolarity condition only via the function
$B(\psi,\Theta,\zeta)$.

As stated in the Introduction, in Boozer coordinates $B_0$ depends
only on a single helicity $M\Theta-N\zeta$. It is enough to carry out
the proof for quasi-axisymmetric $B_0$, i.e. $\partial_\zeta B_0
\equiv 0$, which corresponds to $N=0$. If $\bB_0$ is helically
symmetric, $N\neq 0$, the problem may be reduced to the
quasi-axisymmetric case by a change of Boozer angles. Specifically,
one can define $\overline{\Theta} := M\Theta -N\zeta$ and employ
$\{\psi,\overline{\Theta},\zeta\}$ as Boozer coordinates.

Since $B_1$ can be chosen such that $\langle B_1
\rangle_\psi = 0$ and such that it does not contain the helicity of
$B_0$, we have, in this case,
\begin{equation}\label{eq:conditionB1}
  \int_0^{2\pi} B_1(\psi,\Theta,\zeta)\dd\zeta = 0.
\end{equation}

Now, we are ready to calculate the scaling of
\eq{eq:FSAofradialJdimensionfulBoozer}. Since
\eq{eq:conditionsQuadraticScaling} holds, every term on the right side
of \eq{eq:FSAofradialJdimensionfulBoozer} can be expanded in integer
powers of $\alpha$. The $O(\alpha^0)$ terms vanish due to
quasisymmetry. The $O(\alpha)$ terms, $\left\langle
  \bJ\cdot\nabla\psi \right\rangle_\psi^{(1)}$, are
\begin{eqnarray}\label{eq:J1}
\fl
\left\langle
\bJ\cdot\nabla\psi
\right\rangle_\psi^{(1)}
=
\nonumber\\[5pt]
\fl\hspace{1cm}
-\sum_\spe \frac{m_\spe c}{V'}
\int_0^{2\pi}\int_0^{2\pi}
\dd\Theta\dd\zeta
\int
\frac{u^2 + \mu B_0}{B_0^2}
J \partial_\Theta B_0
\, G_{\spe }^{(1)}
\dd u \dd\mu
\nonumber\\[5pt]
\fl\hspace{1cm}
+\sum_\spe \frac{m_\spe c}{V'}
\int_0^{2\pi}\int_0^{2\pi}
\dd\Theta\dd\zeta
\int
\Bigg[\frac{2u^2 + \mu B_0}{B_0^3}B_1 J\partial_\Theta B_0
\nonumber\\[5pt]
\fl\hspace{1cm}
+
\frac{u^2 + \mu B_0}{B_0^2}
(I\partial_\zeta B_1 - J \partial_\Theta B_1)
\Bigg] G_{\spe }^{(0)}
\dd u \dd\mu,
\end{eqnarray}
where
\begin{equation}\label{eq:expansionOfG1}
G_{\spe 1}:= G_{\spe}^{(0)}+\alpha G_{\spe}^{(1)} +
O(\alpha^2).
\end{equation}
The equations determining $G_{\spe}^{(0)}$ and $ G_{\spe}^{(1)}$
are
\begin{eqnarray}\label{eq:FPG1dimensionfulzero}
\fl
\left(u\bun\cdot\nabla - \mu\bun\cdot\nabla
B\partial_u
\right)^{(0)}
G_{\spe}^{(0)}
\nonumber\\[5pt]
\fl\hspace{1cm}
+\left(
\Upsilon_\spe
v_{\psi,\spe}
 F_{\spe 0}\right)^{(0)}
\nonumber\\[5pt]
\fl\hspace{1cm}
= C^{\ell (0)}_\spe[G^{(0)}]
\end{eqnarray}
and
\begin{eqnarray}\label{eq:FPG1dimensionfulone}
\fl
\left(u\bun\cdot\nabla - \mu\bun\cdot\nabla
B\partial_u
\right)^{(0)}
G_{\spe}^{(1)}
\nonumber\\[5pt]
\fl\hspace{1cm}
+ 
\left(u\bun\cdot\nabla - \mu\bun\cdot\nabla
B\partial_u
\right)^{(1)}
G_{\spe}^{(0)}
\nonumber\\[5pt]
\fl\hspace{1cm}
+\left(
\Upsilon_\spe
v_{\psi,\spe}
 F_{\spe 0}
\right)^{(1)}
\nonumber\\[5pt]
\fl\hspace{1cm}
= C^{\ell(1)}_\spe[G^{(0)}] + C^{\ell(0)}_\spe[G^{(1)}].
\end{eqnarray}
We have employed the notation
\begin{eqnarray}\label{eq:gradparBoozerzero}
\fl  (\bun\cdot\nabla)^{(0)} = 
\frac{2\pi\Psi'_t B_0}{\langle B^2 \rangle_\psi V'}
  (\iotabar\partial_\Theta + \partial_\zeta),
\end{eqnarray}
\begin{eqnarray}\label{eq:gradparBoozerone}
\fl   (\bun\cdot\nabla)^{(1)} = 
\frac{2\pi\Psi'_t B_1}{\langle B^2 \rangle_\psi V'}
  (\iotabar\partial_\Theta + \partial_\zeta),
\end{eqnarray}
\begin{eqnarray}\label{eq:gradparBBoozerzero}
\fl   (\bun\cdot\nabla B)^{(0)} = 
\frac{2\pi\Psi'_t B_0}{\langle B^2 \rangle_\psi V'}
  \iotabar\partial_\Theta B_0,
\end{eqnarray}
\begin{eqnarray}\label{eq:gradparBBoozerone}
\fl   (\bun\cdot\nabla B)^{(1)} = 
\frac{2\pi\Psi'_t}{\langle B^2 \rangle_\psi V'}
\left[
B_1
  \iotabar\partial_\Theta B_0
+
 B_0
  (\iotabar\partial_\Theta + \partial_\zeta) B_1
\right],
\end{eqnarray}
etc. $C^{\ell(0)}_\spe$ is the linearized collision operator
corresponding to $B_0$ and $C^{\ell(1)}_\spe$ is the first-order
correction. Their explicit expressions are not needed here.

From \eq{eq:FPG1dimensionfulzero} and
\eq{eq:FPG1dimensionfulone}, one obtains
\begin{equation}
\partial_\zeta G_\spe^{(0)} = 0
\end{equation}
and
\begin{equation}
\int_0^{2\pi}
G^{(1)}_\spe\dd\zeta = 0.
\end{equation}
Therefore, each term on the right-hand side of \eq{eq:J1} can be
written as
\begin{eqnarray}
\int_0^{2\pi}\int_0^{2\pi}
q(\psi,\Theta) f(\psi,\Theta,\zeta)
\dd\Theta\dd\zeta
\end{eqnarray}
for some function $f(\psi,\Theta,\zeta)$ with zero average over
$\zeta$,
\begin{equation}
\int_0^{2\pi}
f(\psi,\Theta,\zeta)\dd\zeta = 0.
\end{equation}
Then, $\left\langle \bJ\cdot\nabla\psi
\right\rangle_\psi^{(1)}\equiv 0$ follows. The quadratic terms,
$\left\langle \bJ\cdot\nabla\psi \right\rangle_\psi^{(2)}$, are
non-zero in general and we obtain \eq{eq:quadraticScaling}. It is
important to emphasize that the result is valid for any value of the
collisionality. The same scaling was obtained in reference
\cite{Simakov2009} for highly-collisional plasmas.

\section{Breakdown of the $\alpha^2$ scaling}
\label{sec:breakdownOfExpansion}

Clearly, the procedure followed in Section \ref{sec:DeviationsFromQS}
may fail if
\begin{equation}\label{eq:comparable_parallelgradients}
|\alpha\bun_0\cdot\nabla B_1| \sim |\bun_0\cdot\nabla B_0|
\end{equation}
because the parallel streaming operator appearing in the drift kinetic
equation \eq{eq:FPG1dimensionful} cannot be expanded in powers of
$\alpha$. Of course, \eq{eq:comparable_parallelgradients} holds for
any perturbation $\alpha \bB_1$ near points where
$\bun_0\cdot\nabla B_0 = 0$. The trajectories that can be affected
more severely correspond to almost trapped, barely trapped, and deeply
trapped particles in the magnetic field $\bB_0$. However, it has been
proven in \cite{CalvoParraVelascoAlonso13} that they contribute with
terms $O(|\alpha|^{5/2})$ to the radial electric current, and thus are
subdominant with respect to $O(\alpha^2)$ terms. In a sense, these
trajectories, even though they satisfy
\eq{eq:comparable_parallelgradients}, are trivial to treat.

Recall that $L_0$ is the characteristic variation length of $B_0$. We
have to worry about stellarators where
\eq{eq:comparable_parallelgradients} happens due to $|\alpha
\bun_0\cdot\nabla B_1| \sim B_0 L_0^{-1}$, i.e. we are not worried
about the neighborhood of a point with $\bun_0\cdot\nabla B_0 =
0$. We can equivalently say that we have to analyze what happens when
$|\alpha\bun_0\cdot\nabla B_1| \sim |\bun_0\cdot\nabla B_0|$
because the perturbation is such that
\begin{equation}\label{eq:condition_helicity}
\frac{L_1}{L_0} \sim |\alpha|,
\end{equation}
$L_1$ being the characteristic variation length of
$B_1$.

Collisionless particles are expected to be the most dangerous ones,
and we focus on them in this paper by studying the so-called $1/\nu$
regime. In the absence of collisions the kinetic energy $\varepsilon =
u^2/2 + \mu B$ is a constant of the motion, and to lowest order in
$\epsilon_\spe$ particle trajectories lie on magnetic field
lines. This is why it will be useful to employ the phase-space
coordinates $\{\psi,\chi,\Theta,\varepsilon,\mu,s\}$, being $s=-1,1$
the sign of the parallel velocity and $\chi := \Theta -\iotabar\zeta$,
$\chi\in[0,2\pi)$, a coordinate that locally labels magnetic field
lines. Consider the following expansion,
\begin{eqnarray}\label{eq:LowCollExpG1}
  G_{\spe 1} = G_{\spe}^{[-1]} + G_{\spe}^{[0]} 
  + O(\nu_{*\spe}\epsilon_\spe F_{\spe 0}),
\end{eqnarray}
where $G_{\spe}^{[j]} = O(\nu_{*\spe}^{j}\epsilon_\spe F_{\spe
  0})$. Equation \eq{eq:FPG1dimensionful} to lowest order in
$\nu_{*\spe} \ll 1$ reads
\begin{eqnarray}\label{eq:smallcolllowestorder}
v_{||}\bun\cdot\nabla\Theta\partial_\Theta G_\spe^{[-1]} = 0.
\end{eqnarray}
Hence, for passing particles $G_{\spe}^{[-1]}$ is a flux function
whereas for trapped particles $\overline{G_{\spe}^{[-1]}}=
G_{\spe}^{[-1]}$; that is, for trapped particles $G_{\spe}^{[-1]}$ is
not a flux function but it is constant along the lowest order
trajectories. Here,
\begin{equation}\label{eq:DefBounceAverage}
\fl \overline{f} 
= \tau_b^{-1}\oint \frac{f(\psi, \chi, \Theta, \varepsilon, \mu,s) }
{v_{||}\bun\cdot\nabla\Theta}\dd \Theta,
\end{equation}
with
\begin{equation}
\tau_b = \oint \frac{1}
{v_{||}\bun\cdot\nabla\Theta} \dd \Theta,
\end{equation}
defines the bounce average of the phase-space function $f(\psi, \chi,
\Theta, \varepsilon, \mu,s)$. This is a time average over the lowest
order trapped particle trajectories, that are closed, and $\tau_b$ is
the bounce time. The angle $\Theta$ parameterizes the trajectory. In
\eq{eq:DefBounceAverage}, the parallel velocity $v_{||}$ is to be
viewed as a function of the independent variables
$\{\psi,\chi,\Theta,\varepsilon,\mu,s\}$. Namely,
\begin{equation}
  v_{||}(\psi,\chi,\Theta,\varepsilon,\mu,s)
  = s\sqrt{2(\varepsilon - \mu B(\psi,\chi,\Theta))} \, .
\end{equation}

We also point out that if $\{\psi,\Theta,\zeta\}$ are
Boozer coordinates and $\chi = \Theta -\iotabar\zeta$, the flux
surface average of a function $f(\psi,\chi,\Theta)$ reads
\begin{equation}\label{eq:FSAinClebschCoor}
\langle f \rangle_\psi = 
\frac{\langle B^2 \rangle_\psi}{4\pi^2\iotabar}
\int_0^{2\pi}
\int_0^{2\pi}
\frac{1}{B^2}
f(\psi,\chi,\Theta)
\dd\chi\dd\Theta
.
\end{equation}

To an order higher in $\nu_{*\spe}$ than \eq{eq:smallcolllowestorder},
the transit average of the Fokker-Planck equation
\eq{eq:FPG1dimensionful} gives, for trapped particles,
\begin{eqnarray}\label{eq:ConstraintEq1overNuRegime}
\Upsilon_\spe
\overline{v_{\psi,\spe}}
\, F_{\spe 0}
= \overline{C^\ell_\spe[G^{[-1]}]}.
\end{eqnarray}

We proceed to prove that one only has to deal with trapped
trajectories because $G_\spe^{[-1]} = 0$ for passing particles up to
terms $O(\alpha\nu_{*\spe}^{-1}\epsilon_\spe F_{\spe 0})$. Let us denote by
$G_{\spe,p}^{[-1]}$ and $G_{\spe,t}^{[-1]}$ the distribution function
$G_\spe^{[-1]}$ in the passing and trapped regions, respectively.
$G_{\spe,p}^{[-1]}$ is a flux function, whereas
$G_{\spe,t}^{[-1]}$ is constant over orbits. One can write
\begin{equation}
  G_{\spe}^{[-1]}(\psi,\chi,\varepsilon,\mu) 
  = g_{\spe}(\psi,\varepsilon,\mu) 
+ \partial_\chi h_{\spe}(\psi,\chi,\varepsilon,\mu),
\end{equation}
where $h_\spe$ vanishes in the passing region, $h_{\spe,p}\equiv
0$. Observe that by multiplying \eq{eq:FPG1dimensionful} by $g_\spe /
F_{\spe 0}$, integrating over velocity and flux-surface averaging, we
can derive the condition
\begin{eqnarray}\label{eq:ConditionFromDKE}
  \left\langle
\sum_\spe\int \frac{g_{\spe}}{F_{\spe 0}}
C^\ell_\spe[g_{\spe}+ \partial_\chi h_{\spe}]\dd^3 v
\right\rangle_\psi = 0,
\end{eqnarray}
where we have used that for any function $Q(\psi,\varepsilon,\mu)$,
\begin{eqnarray}\label{eq:PropertyRadialMagneticDrift}
\left\langle
\int  v_{\psi,\spe}Q(\psi,\varepsilon,\mu)\dd^3 v
\right\rangle_\psi
=0.
\end{eqnarray}
The proof of \eq{eq:PropertyRadialMagneticDrift} can be found, for
example, in Section 5.1 of reference
\cite{CalvoParraVelascoAlonso13}. Here, we have abbreviated $\int
(\dots) \dd^3v \equiv \sum_s\int_0^\infty\int_0^\infty 2\pi
H(\varepsilon - \mu B) B |v_{||}|^{-1}(\dots)\dd\varepsilon\dd\mu $,
where $H$ is the Heaviside step function.

The kernel of the collision operator in drift-kinetic
coordinates~\cite{CalvoParraVelascoAlonso13} depends on the spatial
coordinates only via $B$, which is simply a function of $\Theta$ up to
terms $O(\alpha)$ because $B_0$ is quasi-axisymmetric. Hence, up to
$O(\alpha)$ corrections, \eq{eq:ConditionFromDKE} becomes
\begin{eqnarray}\label{eq:ConditionFromDKE2}
  \left\langle
\sum_\spe\int \frac{g_{\spe}}{F_{\spe 0}}
C^\ell_\spe[g_{\spe}]\dd^3 v
\right\rangle_\psi = 0,
\end{eqnarray}
where periodicity in $\chi$ has been employed. Equation
\eq{eq:ConditionFromDKE2} can be viewed as an entropy production
condition on $g_\spe$, implying that
\begin{equation}
  g_\spe = (a_{\spe,0}(\bR) + m_\spe a_1(\bR) v_{||} + 
m_\spe a_2(\bR) \varepsilon)F_{\spe 0},
\end{equation}
where we have used that $G_{\spe 1}$ has to be independent of the
gyrophase~\cite{CalvoParraVelascoAlonso13}. Since $g_\spe$ must be a
flux function, $a_{\spe,0} \equiv a_{\spe,0}(\psi)$, $a_{2} \equiv
a_{2}(\psi)$ and $a_{1} \equiv 0$.

One can always choose the flux-surface averaged densities and total
energy to be given only by the $O(\epsilon_\spe^0)$ distribution
function, $F_{\spe 0}$. Thus, we impose
\begin{eqnarray}
\left\langle
\int G_\spe^{[-1]}\dd^3 v
\right\rangle_\psi = 0 \quad \mbox{for every $\spe$}
\end{eqnarray}
and
\begin{eqnarray}
\left\langle
\sum_\spe \int m_\spe \varepsilon G_\spe^{[-1]}\dd^3 v
\right\rangle_\psi = 0.
\end{eqnarray}
To lowest order in $\alpha$, this implies $a_{\spe,0} \equiv 0$ and
$a_{2} \equiv 0$. Then, we deduce that $g_\spe \equiv 0$. Since
$h_\spe$ vanishes in the passing region, we have obtained that
$G_{\spe,p}^{[-1]}\equiv 0$ up to $O(\alpha)$
corrections. Then, to lowest order in $\alpha$,
  passing particles only enter the problem by setting a vanishing
  boundary condition for $G^{[-1]}_{\spe,t}$ at the interface between
  the passing and trapped regions.

Finding out how the solution of equation
\eq{eq:ConstraintEq1overNuRegime} depends on $\alpha$ when $\alpha
B_1$ has a large parallel gradient will be the objective of the
following sections.

\section{Scaling of the bounce-averaged radial magnetic drift in the
  presence of large helicity perturbations}
\label{sec:AverageMagDrift}

As a previous step to finding out how the solution of equation
\eq{eq:ConstraintEq1overNuRegime} scales with $\alpha$ when $L_1\sim
|\alpha| L_0$, we investigate the scaling of
$\overline{v_{\psi,\spe}}$. At the end of Section
\ref{sec:breakdownOfExpansion} we have shown that only trapped
particles require detailed analysis, and consequently we focus on
them. A sketch of a perturbation $\alpha B_1$ with large parallel
gradient is shown in figure \ref{fig:MagneticFieldLine}. We have to
distinguish two cases: a particle trapped in a well of size $L_0$ and
a particle trapped in a secondary well of size $L_1\sim |\alpha|
L_0$. Before starting the analysis of the scaling of
$\overline{v_{\psi,\spe}}$, we remind the reader that along this paper
we assume that the radial magnetic drift can be expanded in integer
powers of $\alpha$. Namely,
\begin{eqnarray}\label{eq:CondMagDriftCanBeExpanded}
v_{\psi,\spe} - v_{\psi,\spe}^{(0)}  =  O(\alpha),
\end{eqnarray}
where $v_{\psi,\spe}^{(0)}$ is the radial magnetic drift corresponding
to $\bB_0$. As argued in the Introduction, equation
\eq{eq:CondMagDriftCanBeExpanded} should be a design criterion for
quasisymmetric stellarators.

\subsection{Large wells}
\label{sec:withoutNewWells}

Take values of energy and magnetic moment such that
  the particle is trapped in a well of size $L_0$ (see the upper red
  line in figure \ref{fig:MagneticFieldLine}). In this subsection we
  will prove that for this particle
\begin{equation}\label{eq:scalingLargeWellAdvance}
\overline{v_{\psi,\spe}} \sim |\alpha|^{1/2}\epsilon_\spe v_{t\spe}
|\nabla\psi|.
\end{equation}
Because of
\eq{eq:CondMagDriftCanBeExpanded}, we may write
\begin{equation}\label{eq:averagemagdriftAux}
  \overline{v_{\psi,\spe}}
  =
  \frac{2\int_{\Theta_1}^{\Theta_2}  v_{\psi,\spe}^{(0)} 
\left[|v_{||}|(\bun\cdot\nabla\Theta)^{(0)}
\right]^{-1} \dd\Theta + O(\alpha)}
  {2\int_{\Theta_1}^{\Theta_2} 
\left[|v_{||}|\bun\cdot\nabla\Theta\right]^{-1}\dd \Theta} \,.
\end{equation}
Here, we denote by $\Theta_1$ and $\Theta_2$ the bounce points of the
orbit; that is, the solutions of $\varepsilon - \mu
B(\psi,\chi,\Theta) =0$. Whereas the denominator of the right side of
\eq{eq:averagemagdriftAux} is $O(L_0/v_{t\spe})$, the integral in the
numerator is dominated by a region near the endpoints whose size is
$O(\alpha)$, that yields the scaling
\eq{eq:scalingLargeWellAdvance}. Next, we proceed to give the proof.

Recalling \eq{eq:vpsiBoozer2}, we get
\begin{equation}\label{eq:vpsizero}
\fl
v_{\psi, \spe}^{(0)}
=
-\frac{2\pi m_\spe c (2\varepsilon-\mu B_0)}
{  Z_\spe e V'\langle B_0^2 \rangle_\psi  B_0}
J \partial_\Theta B_0,
\end{equation}
and using \eq{eq:bdotnablaTheta} we obtain
\begin{equation}
(\bun\cdot\nabla\Theta)^{(0)}
=
\frac{2\pi\Psi'_p B_0}{V'\langle B_0^2\rangle_\psi}.
\end{equation}

The first term in the numerator of \eq{eq:averagemagdriftAux} can then be
recast as
\begin{eqnarray}\label{eq:averagemagdriftAux2}
\fl 2\int_{\Theta_1}^{\Theta_2}  v_{\psi,\spe}^{(0)}
 \left[|v_{||}|(\bun\cdot\nabla\Theta)^{(0)}\right]^{-1}
 \dd \Theta =
\nonumber\\[5pt]
\fl \hspace{1cm}
-\frac{ 2 m_\spe c J}
{  Z_\spe e \Psi'_p  }\int_{\Theta_1}^{\Theta_2}
\frac{ \partial_\Theta B_0(\psi,\Theta)}
{ B_0^2(\psi,\Theta)}
\frac{2\varepsilon-\mu B_0(\psi,\Theta)}
{\sqrt{2(\varepsilon - \mu B_0(\psi,\Theta)-\mu\alpha B_1(\psi,\chi,\Theta)}}
 \dd \Theta
.
\end{eqnarray}
To simplify the notation, in what follows either we will omit the
arguments of $B_0$ and $B_1$ or we will only specify the dependence on
$\Theta$.

The identity
\begin{eqnarray}\label{eq:propertyZeroAveragedDrfitVelocity}
-\frac{2 m_\spe c J}{Z_\spe e \Psi'_p}\int_{\Theta_1}^{\Theta_2}
\frac{\partial_\Theta \tilde B}{\tilde B^2}
 \frac{2\varepsilon - \mu \tilde B}
{\sqrt{2(\varepsilon - \mu \tilde B)}} \dd \Theta
= 0
\end{eqnarray}
with
\begin{equation}
\tilde B(\Theta) := B_0(\Theta) +\alpha
B_1(\Theta_2)\frac{\Theta-\Theta_1}{\Theta_2-\Theta_1}
+\alpha
B_1(\Theta_1)\frac{\Theta_2-\Theta}{\Theta_2-\Theta_1}
\end{equation}
allows us to rewrite \eq{eq:averagemagdriftAux2} as
\begin{eqnarray}\label{eq:averagemagdriftAux3}
\fl 2\int_{\Theta_1}^{\Theta_2}  v_{\psi,\spe}^{(0)}
 \left[|v_{||}|(\bun\cdot\nabla\Theta)^{(0)}\right]^{-1}
 \dd \Theta =
\nonumber\\[5pt]
\fl \hspace{1cm}
-2\frac{ m_\spe c J}
{  Z_\spe e \Psi'_p  }
\int_{\Theta_1}^{\Theta_2}
\Bigg(
\frac{ \partial_\Theta B_0}
{ B_0^2}
\frac{2\varepsilon-\mu B_0}
{\sqrt{2(\varepsilon - \mu B_0-\mu\alpha B_1)}}
\nonumber\\[5pt]
\fl \hspace{1cm}
-
\frac{ \partial_\Theta \tilde B}
{ \tilde B^2}
\frac{2\varepsilon-\mu \tilde B}
{\sqrt{2(\varepsilon - \mu \tilde B)}}
\Bigg)
 \dd \Theta.
\end{eqnarray}
We want to prove that this integral is dominated by a neighborhood of
the endpoints $\Theta_1$ and $\Theta_2$ and that it scales with
$|\alpha|^{1/2}$.

Take $K > 1$ and choose $\Theta_1'$ and $\Theta_2'$,
$[\Theta_1',\Theta_2']\subset[\Theta_1,\Theta_2]$, such that
$\varepsilon - \mu B_0(\Theta) >|\alpha| K$ when
$\Theta\in[\Theta_1',\Theta_2']$. It is convenient to select
$\Theta_1'$ and $\Theta_2'$ such that $\tilde B(\Theta_1') = \tilde
B(\Theta_2')$.  First, we show that in \eq{eq:averagemagdriftAux3} the
piece of the integral that runs over $[\Theta_1',\Theta_2']$ is
negligible compared to $|\alpha|^{1/2}$.

Using that $\tilde B(\Theta_1') = \tilde
B(\Theta_2')$, we have
\begin{eqnarray}\label{eq:propertyZeroAveragedDrfitVelocityPrimes}
-\frac{2 m_\spe c J}{Z_\spe e \Psi'_p}\int_{\Theta_1'}^{\Theta_2'}
\frac{\partial_\Theta \tilde B}{\tilde B^2}
 \frac{2\varepsilon - \mu \tilde B}
{\sqrt{2(\varepsilon - \mu \tilde B)}} \dd \Theta
= 0.
\end{eqnarray}
Proving that the integration over $[\Theta_1',\Theta_2']$ in the first
term of \eq{eq:averagemagdriftAux3} is small requires
some work. To fix ideas, assume that
\begin{eqnarray}\label{eq:IntThetasWithPrime}
\int_{\Theta_1'}^{\Theta_2'}
\frac{ \partial_\Theta B_0}
{ B_0^2}
\frac{2\varepsilon-\mu B_0}
{\sqrt{2(\varepsilon - \mu B_0-\mu\alpha B_1)}}
\dd\Theta
\end{eqnarray}
is positive (if it is negative, the treatment is almost identical). Then,
\begin{eqnarray}\label{eq:IntThetasWithPrime2}
\fl
\int_{\Theta_1'}^{\Theta_2'}
\frac{ \partial_\Theta B_0}
{ B_0^2}
\frac{2\varepsilon-\mu B_0}
{\sqrt{2(\varepsilon - \mu B_0-\mu\alpha B_1)}}
\dd\Theta
\le
\nonumber\\[5pt]
\fl\hspace{1cm}
\int_{\Theta_1'}^{\Theta_m}
\frac{ \partial_\Theta B_0}
{ B_0^2}
\frac{2\varepsilon-\mu B_0}
{\sqrt{2(\varepsilon - \mu B_0-\mu\alpha B_{1,min})}}
\dd\Theta
\nonumber\\[5pt]
\fl\hspace{1cm}
+
\int_{\Theta_m}^{\Theta_2'}
\frac{ \partial_\Theta B_0}
{ B_0^2}
\frac{2\varepsilon-\mu B_0}
{\sqrt{2(\varepsilon - \mu B_0-\mu\alpha B_{1,max})}}
\dd\Theta,
\end{eqnarray}
where $\partial_\Theta B_0(\Theta_m)=0$. Therefore, $\partial_\Theta
B_0(\Theta)<0$ for $\Theta\in[\Theta_1',\Theta_m)$ and
$\partial_\Theta B_0(\Theta)>0$ for
$\Theta\in(\Theta_m,\Theta_2']$. The minimum of $B_1(\Theta)$ in
$[\Theta_1',\Theta_m)$ has been denoted by $B_{1,min}$ and the maximum
of $B_1(\Theta)$ in $(\Theta_m,\Theta_2']$ by $B_{1,max}$. Note that
by performing these integrals in $[\Theta_1',\Theta_2']$ we have been
able to give bounds for the integrand, that tends to infinity at
$\Theta = \Theta_1$ and $\Theta = \Theta_2$.

Let us manipulate the first term on the right side of
\eq{eq:IntThetasWithPrime2}. Trivially,
\begin{eqnarray}\label{eq:IntThetasWithPrime3}
\fl
\int_{\Theta_1'}^{\Theta_m}
\frac{ \partial_\Theta B_0}
{ B_0^2}
\frac{2\varepsilon-\mu B_0}
{\sqrt{2(\varepsilon - \mu B_0-\mu\alpha B_{1,min})}}
\dd\Theta =
\nonumber\\[5pt]
\fl\hspace{1cm}
\int_{\Theta_1'}^{\Theta_m}
\frac{ \partial_\Theta B_0}
{ B_0^2}
\frac{2\varepsilon - 2\mu\alpha B_{1,min}-\mu B_0}
{\sqrt{2(\varepsilon - \mu B_0-\mu\alpha B_{1,min})}}
\dd\Theta
\nonumber\\[5pt]
\fl\hspace{1cm}
+
\int_{\Theta_1'}^{\Theta_m}
\frac{ \partial_\Theta B_0}
{ B_0^2}
\frac{2\mu\alpha B_{1,min}}
{\sqrt{2(\varepsilon - \mu B_0-\mu\alpha B_{1,min})}}
\dd\Theta.
\end{eqnarray}
The integrand of the first term on the right side is an exact
differential, whereas the second term is expressed in a useful way
after an integration by parts. The result is
\begin{eqnarray}\label{eq:IntThetasWithPrime4}
\fl
\int_{\Theta_1'}^{\Theta_m}
\frac{ \partial_\Theta B_0}
{ B_0^2}
\frac{2\varepsilon-\mu B_0}
{\sqrt{2(\varepsilon - \mu B_0-\mu\alpha B_{1,min})}}
\dd\Theta =
\nonumber\\[5pt]
\fl\hspace{1cm}
=
-\frac{1}{B_0(\Theta_m)}
\sqrt{2(\varepsilon - \mu B_0(\Theta_m)-\mu\alpha B_{1,min})}
\nonumber\\[5pt]
\fl\hspace{1cm}
+
\frac{1}{B_0(\Theta_1')}
\sqrt{2(\varepsilon - \mu B_0(\Theta_1')-\mu\alpha B_{1,min})}
\nonumber\\[5pt]
\fl\hspace{1cm}
-
\frac{2\alpha B_{1,min}}{B_0^2(\Theta_m)}
\sqrt{2(\varepsilon - \mu B_0(\Theta_m)-\mu\alpha B_{1,min})}
\nonumber\\[5pt]
\fl\hspace{1cm}
+
\frac{2\alpha B_{1,min}}{B_0^2(\Theta_1')}
\sqrt{2(\varepsilon - \mu B_0(\Theta_1')-\mu\alpha B_{1,min})}
\nonumber\\[5pt]
\fl\hspace{1cm}
+
\int_{\Theta_1'}^{\Theta_m}
\sqrt{2(\varepsilon - \mu B_0(\Theta)-\mu\alpha B_{1,min})}
\, \partial_\Theta\left(\frac{\alpha B_{1,min} }{B_0^2}\right)
\dd\Theta
.
\end{eqnarray}
The three last terms (and hence the second term on the right side of
\eq{eq:IntThetasWithPrime3}) are clearly $O(\alpha)$. Analogous
manipulations on the last term of \eq{eq:IntThetasWithPrime2} give
\begin{eqnarray}\label{eq:IntThetasWithPrime5}
\fl
\int_{\Theta_1'}^{\Theta_2'}
\frac{ \partial_\Theta B_0}
{ B_0^2}
\frac{2\varepsilon-\mu B_0}
{\sqrt{2(\varepsilon - \mu B_0-\mu\alpha B_1)}}
\dd\Theta
\le
\nonumber\\[5pt]
\fl\hspace{1cm}
\frac{1}{B_0(\Theta_m)}
\sqrt{2(\varepsilon - \mu B_0(\Theta_m)-\mu\alpha B_{1,max})}
\nonumber\\[5pt]
\fl\hspace{1cm}
-\frac{1}{B_0(\Theta_m)}
\sqrt{2(\varepsilon - \mu B_0(\Theta_m)-\mu\alpha B_{1,min})}
\nonumber\\[5pt]
\fl\hspace{1cm}
+
\frac{1}{B_0(\Theta_1')}
\sqrt{2(\varepsilon - \mu B_0(\Theta_1')-\mu\alpha B_{1,min})}
\nonumber\\[5pt]
\fl\hspace{1cm}
-
\frac{1}{B_0(\Theta_2')}
\sqrt{2(\varepsilon - \mu B_0(\Theta_2')-\mu\alpha B_{1,max})}
=
O\left(\sqrt{\frac{|\alpha|}{K}}\right).
\end{eqnarray}
To write the last equality we have employed that the combination of
the first two terms on the right side of \eq{eq:IntThetasWithPrime5}
is $O(\alpha)$. As for the last two terms, we have used that
$\varepsilon - \mu B_0(\Theta) >|\alpha| K$ when
$\Theta\in[\Theta_1',\Theta_2']$ and that $B_0(\Theta_2') -
B_0(\Theta_1') = O(\alpha)$. The latter is an immediate consequence of
$\tilde B(\Theta_1') = \tilde B(\Theta_2')$. Thus,
\begin{eqnarray}
\fl
\frac{1}{B_0(\Theta_1')}
\sqrt{2(\varepsilon - \mu B_0(\Theta_1')-\mu\alpha B_{1,min})}
\nonumber\\[5pt]
\fl\hspace{1cm}
-
\frac{1}{B_0(\Theta_2')}
\sqrt{2(\varepsilon - \mu B_0(\Theta_2')-\mu\alpha B_{1,max})}
=
O\left(\sqrt{\frac{|\alpha|}{K}}\right).
\end{eqnarray}

Hence, we have shown that
\begin{eqnarray}\label{eq:averagemagdriftWithPrimes}
\fl 2\int_{\Theta_1}^{\Theta_2}  v_{\psi,\spe}^{(0)}
 \left[|v_{||}|(\bun\cdot\nabla\Theta)^{(0)}\right]^{-1}
 \dd \Theta =
\nonumber\\[5pt]
\fl \hspace{1cm}
-2\frac{ m_\spe c J}
{  Z_\spe e \Psi'_p  }
\int_{\Theta_1}^{\Theta_1'}
\Bigg(
\frac{ \partial_\Theta B_0}
{ B_0^2}
\frac{2\varepsilon-\mu B_0}
{\sqrt{2(\varepsilon - \mu B_0-\mu\alpha B_1)}}
\nonumber\\[5pt]
\fl \hspace{1cm}
-
\frac{ \partial_\Theta \tilde B}
{ \tilde B^2}
\frac{2\varepsilon-\mu \tilde B}
{\sqrt{2(\varepsilon - \mu \tilde B)}}
\Bigg)
 \dd \Theta
+
\nonumber\\[5pt]
\fl \hspace{1cm}
-2\frac{ m_\spe c J}
{  Z_\spe e \Psi'_p  }
\int_{\Theta_2'}^{\Theta_2}
\Bigg(
\frac{ \partial_\Theta B_0}
{ B_0^2}
\frac{2\varepsilon-\mu B_0}
{\sqrt{2(\varepsilon - \mu B_0-\mu\alpha B_1)}}
\nonumber\\[5pt]
\fl \hspace{1cm}
-
\frac{ \partial_\Theta \tilde B}
{ \tilde B^2}
\frac{2\varepsilon-\mu \tilde B}
{\sqrt{2(\varepsilon - \mu \tilde B)}}
\Bigg)
 \dd \Theta
+
O\left(\sqrt{\frac{|\alpha|}{K}}\right)
.
\end{eqnarray}
Using that $\Theta_1'-\Theta_1 = O(\alpha K)$ and expanding
$B_0(\Theta) = B_0(\Theta_1) + \partial_\Theta B_0(\Theta_1)(\Theta-\Theta_1) +
O((\Theta-\Theta_1)^2)$, it is easy to demonstrate that
\begin{eqnarray}\label{eq:averagemagdriftWithPrimesAux}
\fl
\int_{\Theta_1}^{\Theta_1'}
\Bigg(
\frac{ \partial_\Theta B_0}
{ B_0^2}
\frac{2\varepsilon-\mu B_0}
{\sqrt{2(\varepsilon - \mu B_0-\mu\alpha B_1)}}
\nonumber\\[5pt]
\fl \hspace{1cm}
-
\frac{ \partial_\Theta \tilde B}
{ \tilde B^2}
\frac{2\varepsilon-\mu \tilde B}
{\sqrt{2(\varepsilon - \mu \tilde B)}}
\Bigg)
 \dd \Theta
=
\nonumber\\[5pt]
\fl \hspace{1cm}
\frac{ \partial_\Theta B_0(\Theta_1)}
{ B_0^2(\Theta_1)}
(2\varepsilon-\mu B_0(\Theta_1))
\times
\nonumber\\[5pt]
\fl \hspace{1cm}
\int_{0}^{\Theta_1'-\Theta_1}
\Bigg(
\frac{1}
{\sqrt{2[- \mu \partial_\Theta B_0(\Theta_1)\Delta
-\mu\alpha (B_1(\Theta_1 + \Delta)-B_1(\Theta_1))]}}
\nonumber\\[5pt]
\fl \hspace{1cm}
-
\frac{1}
{\sqrt{-2 \mu \partial_\Theta B_0(\Theta_1)\Delta
}}\Bigg)\dd\Delta
+
O(\alpha K),
\end{eqnarray}
where the change of variable $\Delta := \Theta - \Theta_1$ has been
performed. Noting that $B_1$ can be extended to
  $\Theta\in(-\infty,\infty)$ without problem and Taylor expanding in
  $\alpha/\Delta$, one gets
\begin{eqnarray}
\fl
\int_{\Theta_1'-\Theta_1}^\infty
\Bigg(
\frac{1}
{\sqrt{2[- \mu \partial_\Theta B_0(\Theta_1)\Delta
-\mu\alpha (B_1(\Theta_1 + \Delta)-B_1(\Theta_1))]}}
\nonumber\\[5pt]
\fl \hspace{1cm}
-
\frac{1}
{\sqrt{-2 \mu \partial_\Theta B_0(\Theta_1)\Delta
}}\Bigg)\dd\Delta \sim
\nonumber\\[5pt]
\fl \hspace{1cm}
\int_{\Theta_1'-\Theta_1}^\infty\frac{\alpha}{\Delta^{3/2}}\dd\Delta
=
O\left(\sqrt{\frac{|\alpha|}{K}}\,\right).
\end{eqnarray}
Similar considerations for the integral over
$[\Theta_2',\Theta_2]$ in \eq{eq:averagemagdriftWithPrimes} yield,
finally,
\begin{eqnarray}\label{eq:averagemagdriftWithPrimesAux2}
\fl 2\int_{\Theta_1}^{\Theta_2}  v_{\psi,\spe}^{(0)}
 \left[|v_{||}|(\bun\cdot\nabla\Theta)^{(0)}\right]^{-1}
 \dd \Theta =
\nonumber\\[5pt]
\fl \hspace{1cm}
-2\frac{ m_\spe c J}
{  Z_\spe e \Psi'_p  }
\frac{ \partial_\Theta B_0(\Theta_1)}
{ B_0^2(\Theta_1)}
(2\varepsilon-\mu B_0(\Theta_1))
\times
\nonumber\\[5pt]
\fl \hspace{1cm}
\int_{0}^{\infty}
\Bigg(
\frac{1}
{\sqrt{2[- \mu \partial_\Theta B_0(\Theta_1)\Delta
-\mu\alpha (B_1(\Theta_1+\Delta)-B_1(\Theta_1))]}}
\nonumber\\[5pt]
\fl \hspace{1cm}
-
\frac{1}
{\sqrt{-2 \mu \partial_\Theta B_0(\Theta_1)\Delta
}}\Bigg)\dd\Delta
\nonumber\\[5pt]
\fl \hspace{1cm}
-2\frac{ m_\spe c J}
{  Z_\spe e \Psi'_p  }
\frac{ \partial_\Theta B_0(\Theta_2)}
{ B_0^2(\Theta_2)}
(2\varepsilon-\mu B_0(\Theta_2))
\times
\nonumber\\[5pt]
\fl \hspace{1cm}
\int_{0}^{\infty}
\Bigg(
\frac{1}
{\sqrt{2[\mu \partial_\Theta B_0(\Theta_2)\Delta
-\mu\alpha (B_1(\Theta_2-\Delta)-B_1(\Theta_2))]}}
\nonumber\\[5pt]
\fl \hspace{1cm}
-
\frac{1}
{\sqrt{2 \mu \partial_\Theta B_0(\Theta_2)\Delta
}}\Bigg)\dd\Delta
+
O\left(\sqrt{\frac{|\alpha|}{K}}\right) + O(\alpha K)
.
\end{eqnarray}
By choosing $K \sim |\alpha|^{-1/3}$, the error is
  minimized and the resulting corrections in
  \eq{eq:averagemagdriftWithPrimesAux2} are $O(|\alpha|^{2/3})$.
  Written this way, it is manifest that if $\partial_\Theta
  B_1(\Theta) \sim \alpha^{-1} B_1(\Theta)/L_0$ as assumed in this
  section, then the right-hand side of
  \eq{eq:averagemagdriftWithPrimesAux2}, and therefore the numerator
  of \eq{eq:averagemagdriftAux}, scales with $|\alpha|^{1/2}$. For the
  bounce time we have
\begin{equation}
\tau_b = 2\int_{\Theta_1}^{\Theta_2} 
\left[|v_{||}|\bun\cdot\nabla\Theta\right]^{-1}\dd \Theta
\sim \frac{L_0}{v_{t\spe}},
\end{equation}
so that the bounce-averaged radial magnetic drift of particles trapped in
large wells exhibits the scaling announced in
\eq{eq:scalingLargeWellAdvance}.

We point out that particles trapped in a large well, but passing
sufficiently close to a new X point created by the perturbation,
present some peculiarities because the bounce time may be arbitrarily
large. Since there are very few of them, we discuss the topic in
subsection \ref{sec:withNewWells}. Particles barely trapped in
secondary wells exhibit the same phenomenon and represent a
significant fraction of all particles trapped in such wells.

\subsection{Secondary wells}
\label{sec:withNewWells}

\begin{figure}
\includegraphics[width=\columnwidth]{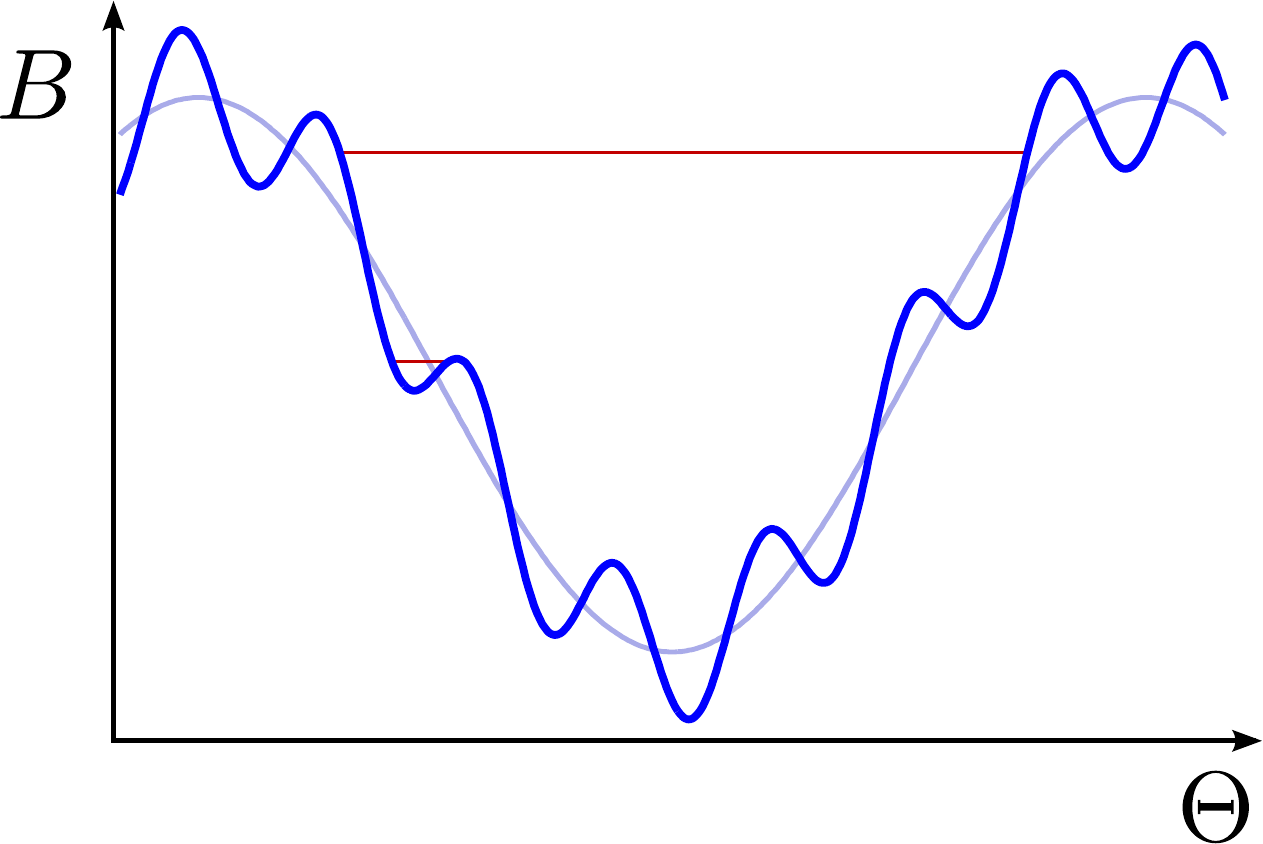}
\caption{Dependence of the magnetic field magnitude
    on $\Theta$ along a magnetic field line in a quasisymmetric
    configuration (light) and in a quasisymmetric configuration with a
    large helicity perturbation added (dark). The upper red line
    corresponds to the trajectory of a particle trapped in a well of
    size $L_0$. The lower one corresponds to a particle trapped in a
    secondary well of size $L_1$.}
\label{fig:MagneticFieldLine}
\end{figure}

Consider a particle trapped in one of the small
  secondary wells of figure \ref{fig:MagneticFieldLine}. The size of
  the new wells is $L_1\sim |\alpha| L_0$ and the characteristic
  parallel velocity of particles trapped in them is $v_{||}\sim
  |\alpha|^{1/2}v_{t\spe}$. Then, it is straightforward to realize
  that
\begin{equation}\label{eq:ScalingaveragemagdriftSecWell}
  \overline{v_{\psi,\spe}}\approx v_{\psi,\spe}^{(0)}(\Theta_{1,0})
\sim \epsilon_\spe v_{t\spe} |\nabla\psi|,
\end{equation}
where $\Theta_{1,0}$ is one of the bounce points for $\bB_0$, i.e. a
solution of $\varepsilon-\mu B_0(\Theta_{1,0}) = 0$. To find
\eq{eq:ScalingaveragemagdriftSecWell} we have Taylor expanded
$v_{\psi,\spe}^{(0)}(\Theta)$ around $\Theta_{1,0}$.

Next, we comment on a subtle point. Whereas
\eq{eq:ScalingaveragemagdriftSecWell} is correct for all particles
trapped in secondary wells, these particles can be split into two
families as the scaling with $\alpha$ of their bounce time is
concerned. For a typical particle trapped in a secondary well,
\begin{eqnarray}
  \oint \frac{v_{\psi,\spe}^{(0)}}{v_{||}\bun\cdot\nabla\Theta}\dd\Theta 
\sim 
|\alpha|^{1/2}\epsilon_\spe L_0 |\nabla\psi|
\end{eqnarray}
and
\begin{eqnarray}
  \tau_b = \oint \frac{1}{v_{||}\bun\cdot\nabla\Theta}\dd\Theta
\sim \frac{L_1}{|\alpha|^{1/2}v_{t\spe}} \sim \frac{|\alpha|^{1/2} L_0}{v_{t\spe}},
\end{eqnarray}
which, of course, yield \eq{eq:ScalingaveragemagdriftSecWell}. But
there is another interesting type of trajectories, whose discussion is
more convenient in coordinates $\{\bR,\varepsilon,\lambda,s\}$, where
\begin{equation}
\lambda := \mu
{\cal B}_0 /\varepsilon
\end{equation}
is the pitch-angle and ${\cal
  B}_0:= \sqrt{\langle B^2\rangle_\psi}$. In these coordinates, the
parallel velocity reads
\begin{equation}\label{eq:vparpitchangle}
v_{||} = s\sqrt{2\varepsilon}\sqrt{1-\lambda\frac{B}{{\cal B}_0}}\,.
\end{equation}

\begin{figure}
\includegraphics[width=1\columnwidth]{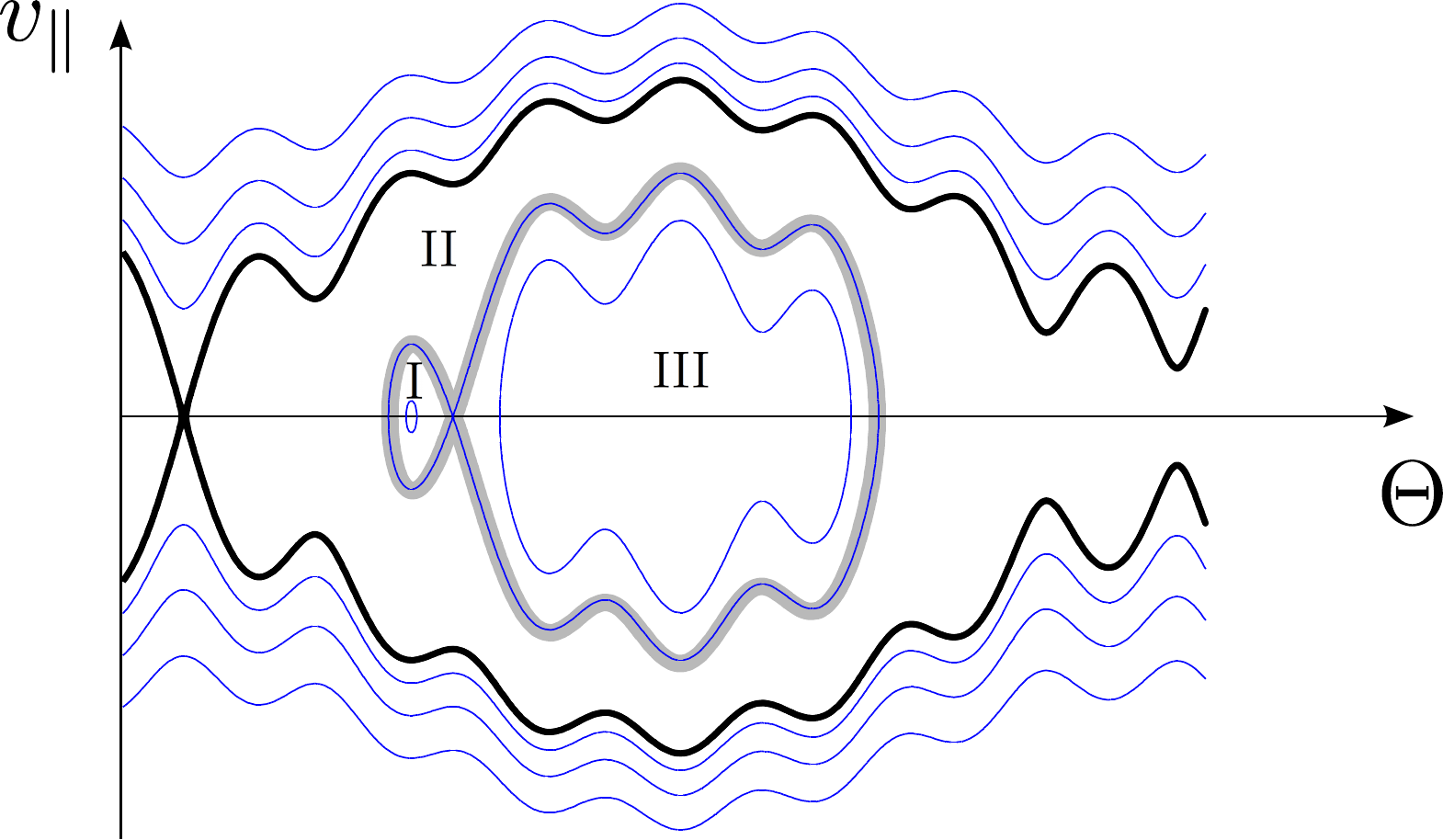}
\caption{Contours of constant kinetic energy and magnetic moment for
  the perturbed quasisymmetric field of figure
  \ref{fig:MagneticFieldLine}. The regions into which phase space is
  divided for the calculation of Section \ref{sec:Matching} are
  shown. Regions I, II and III are labeled. Region IV is the gray
  area surrounding Regions I and III.}
\label{fig:EnergyContours}
\end{figure}

The perturbation $\bB_1$ has created new X points in phase space, one
of which is clearly seen in the gray area of figure
\ref{fig:EnergyContours}. Define $\lambda_M = {\cal B}_0/B_M$, where
$B_M = B(\Theta_M)$ and $\Theta_M$ is the value of $\Theta$ at which
the X point is located. Since it corresponds to a local maximum of
$B(\Theta)$, it is not difficult realize that trajectories with
$\lambda = \lambda_M$ present a logarithmic divergence in
\begin{eqnarray}
  \oint \frac{v_{\psi,\spe}^{(0)}}{v_{||}\bun\cdot\nabla\Theta}\dd\Theta
\end{eqnarray}
and in the integral that gives the bounce time,
\begin{eqnarray}
  \tau_b = \oint \frac{1}{v_{||}\bun\cdot\nabla\Theta}\dd\Theta.
\end{eqnarray}
Let us be more precise. If we define $\delta\lambda := \lambda -
\lambda_M$, then, for a particle trapped in the secondary well,
$\delta\lambda$ is positive. And if $\delta\lambda\lesssim|\alpha|$,
one has
\begin{eqnarray}\label{eq:intlogdiv}
  \oint \frac{v_{\psi,\spe}^{(0)}}{v_{||}\bun\cdot\nabla\Theta}\dd\Theta 
\sim |\alpha|^{1/2}\ln\left|\frac{\alpha}{\delta\lambda}\right| \epsilon_\spe L_0 |\nabla\psi|
\end{eqnarray}
and
\begin{eqnarray}\label{eq:intlogdivtau}
  \tau_b \sim |\alpha|^{1/2}\ln\left|\frac{\alpha}{\delta\lambda}\right|
\frac{L_0}{v_{t\spe}}.
\end{eqnarray}
It is not difficult to derive these scalings by noting that
$\partial^2_\Theta B(\Theta) \sim \alpha^{-1} B_0$, that in a
neighborhood of $\lambda_M$ and $\Theta_M$ the expression
\eq{eq:vparpitchangle} for the parallel velocity can be approximated
by
\begin{equation}
  v_{||} \approx s\sqrt{\frac{2\varepsilon B_M}{{\cal B}_0}}
\sqrt{\frac{\lambda_M|\partial_\Theta^2 B(\Theta_M)|}{2B_M}
(\Theta-\Theta_M)^2
-\delta\lambda}\,,
\end{equation}
and by observing that the integrals involved in the computation of
\eq{eq:intlogdiv} and \eq{eq:intlogdivtau} are dominated by a region
of size $O(\alpha)$ in $\Theta$.

Physically, this result means that particles with small
$\delta\lambda$ have large bounce times and, for $\delta\lambda$
strictly equal to $0$, the particle never reaches the bounce point at
$\Theta = \Theta_M$. However, the ratio of \eq{eq:intlogdiv} and
\eq{eq:intlogdivtau} is such that
\eq{eq:ScalingaveragemagdriftSecWell} is satisfied. Anyway, the
logarithmic corrections do not affect the size of the distribution
function given in Section \ref{sec:Matching}, and the same is true for
particles almost trapped in the secondary well briefly mentioned at
the end of subsection \ref{sec:withoutNewWells}.

\section{Scaling of the distribution function and the flux-surface
  averaged radial electric current}
\label{sec:Matching}

We have seen that $\overline{v_{\psi,\spe}}$ scales differently with
$\alpha$ depending on whether the particle is trapped in a large well
or trapped in a small well. In order to determine the asymptotic
behavior with $\alpha$ of $G_{\spe}^{[-1]}$, we need to solve the
problem independently in several regions of phase space, these regions
differing mostly on their characteristic values for $v_{||}$. Then,
global properties will impose matching conditions on the different
pieces of $G_{\spe}^{[-1]}$. Specifically, we divide the phase space
in four regions that are shown in figure \ref{fig:EnergyContours} and
will be described in more detail below.

From now on, we assume that the plasma consists of electrons and
singly-charged ions, and solve for the ions. Then, the equation to be
solved for trapped trajectories is
\begin{eqnarray}\label{eq:DKEionsOrbitAveraged}
\Upsilon_\spe \overline{v_{\psi,i}}F_{i 0} = 
\overline{C_{ii}^\ell[G_i^{[-1]}]},
\end{eqnarray}
where we have employed that the ion-electron collision term is small
by $\sqrt{m_e/m_i}\ll 1$. A remark about our assumptions on the value
of the collisionality is in order here. Equation
\eq{eq:DKEionsOrbitAveraged} is correct for trapped particles whose
bounce frequency $\tau_b^{-1}$ is much larger than their effective
collision frequency $\nu_{{\rm eff}}$. A particle trapped in a large
well of size $L_0$ has $v_{||} \sim v_{ti}$, $\tau_b^{-1}\sim
v_{ti}/L_0$ and $\nu_{{\rm eff}}\sim \nu_i$. Hence, for them,
$\nu_{{\rm eff}}\ll \tau_b^{-1}$ is equivalent to
\begin{equation}\label{eq:LowCollCond}
\frac{\nu_{i}L_0}{v_{ti}}\ll 1,
\end{equation}
which is what one usually understands by low collisionality
regime. However, a particle trapped in a secondary well of size
$L_1\sim |\alpha| L_0$ has a parallel velocity $v_{||}\sim
|\alpha|^{1/2}v_{ti}$, a bounce frequency
$\tau_b^{-1} \sim |\alpha|^{-1/2}v_{ti}/L_0$ and an effective collision
frequency $\nu_{{\rm eff}}\sim \nu_i/\alpha$. Then, the condition of
small collisionality for these particles amounts to requiring
\begin{equation}\label{eq:LowCollCondSmallWell}
  \frac{\nu_{i}L_0}{\alpha^{1/2}v_{ti}}\ll 1,
\end{equation}
which is more demanding that \eq{eq:LowCollCond}. We assume that both
\eq{eq:LowCollCond} and \eq{eq:LowCollCondSmallWell} are satisfied so
that \eq{eq:DKEionsOrbitAveraged} is the equation determining the
distribution function for all trapped trajectories, i.e. all trapped
particles are in a $1/\nu$ regime (except for the collisional layer
shown in figure \ref{fig:EnergyContours}).

Let us start by determining the distribution function in the region
corresponding to a small, secondary well, denoted by Region I. Its
size is of order $L_1\sim |\alpha| L_0$ and the associated parallel
velocities are $v_{||}\sim|\alpha|^{1/2}v_{t\spe}$. The pitch-angle
scattering term dominates in the collision operator of the right-hand
side of \eq{eq:DKEionsOrbitAveraged} because $v_{||}$ and $\mu$ have
very different scales, leaving us with
\begin{eqnarray}\label{eq:DKEwell}
 \Upsilon_\spe\overline{v_{\psi,i}}F_{i 0} = 
\overline{
\frac{v_{||}}{2\varepsilon \tau_\perp}\frac{{\cal B}_0}{B}\partial_\lambda
\left(v_{||}\lambda\partial_\lambda G_i^I\right)
}\, .
\end{eqnarray}
Here, $G_i^{\mathrm{I}}$ stands for the lowest order piece in $\alpha$ of
$G_i^{[-1]}$ in Region I. The explicit expression for the perpendicular
collisional time is
\begin{eqnarray}
\frac{1}{\tau_\perp(\varepsilon)}
=
\frac{8\pi e^4 n_i \ln\Lambda}{m_i^2(2\varepsilon)^{3/2}}
[\Phi(\sqrt{m_i \varepsilon /T_i}) - \beta(\sqrt{m_i \varepsilon /T_i})],
\end{eqnarray}
where $\ln\Lambda$ is the Coulomb logarithm,
\begin{equation}
\Phi(x) = \frac{2}{\sqrt{\pi}}\int_0^x e^{-y^2}\dd y
\end{equation}
and
\begin{equation}
\beta(x) = \frac{\Phi(x)-x\Phi'(x)}{2x^2}.
\end{equation}

From \eq{eq:DKEwell} one immediately obtains
\begin{eqnarray}\label{eq:DKEwell2}
\fl
\int_{\Theta_1}^{\Theta_2} \frac{ \Upsilon_\spe v_{\psi,i}}{|v_{||}|\bun\cdot\nabla\Theta}
  F_{i 0} \dd\Theta = 
  \int_{\Theta_1}^{\Theta_2}
  \frac{{\cal B}_0}{2\varepsilon \tau_\perp \bB\cdot\nabla\Theta}
\partial_\lambda \left(|v_{||}|\lambda\partial_\lambda G_i^{\mathrm{I}}\right)
\dd \Theta
,
\end{eqnarray}
$\Theta_1$ and $\Theta_2$ being the bounce points. Hence,
\begin{eqnarray}\label{eq:DKEwell3}
\fl
\partial_\lambda G_i^{\mathrm{I}}
=
\left(\int_{\Theta_1}^{\Theta_2}
\frac{{\cal B}_0}{2\varepsilon \tau_\perp \bB\cdot\nabla\Theta}
|v_{||}|
\dd \Theta
\right)^{-1}\frac{1}{\lambda}\int_{\lambda_b}^\lambda
\int_{\Theta_1}^{\Theta_2} \frac{\Upsilon_\spe v_{\psi,i}}{|v_{||}|\bun\cdot\nabla\Theta}
F_{i 0} \dd\Theta,
\end{eqnarray}
where we have used
  that $G_i^{\mathrm{I}}$ is constant over the orbit and we have assumed the 
regularity condition $v_{||}
\partial_\lambda G^{\mathrm{I}}_i\vert_{\lambda_b}=0$ at the bottom of
the well, $\lambda_b = {\cal B}_0/B_b$, with $B_b$
  the minimum value of $B$ in the well. Observe that \eq{eq:DKEwell3}
completely determines $\partial_\lambda G^{\mathrm{I}}_i$ inside the
secondary well. Finally, noting that the size of the well is
$O(\alpha)$ in the pitch-angle coordinate, one learns that
\begin{equation}
\partial_\lambda G_i^{\mathrm{I}} \sim 
\nu_{*i}^{-1}\epsilon_i F_{i0}.
\end{equation}
It is easy to convince oneself that the logarithmic corrections in
\eq{eq:intlogdiv} associated to particles barely trapped in the
secondary well give subdominant contributions after performing the
integral over $\lambda$ in equation \eq{eq:DKEwell3}.

We have obtained $\partial_\lambda G_i^{\mathrm{I}}$, but we do not know the size
of $G_i^{\mathrm{I}}$ yet. For this, we need to know the distribution function in
Regions II and III and integrate in $\lambda$ from Regions II and III
towards $\lambda_b$.

In Regions II and III one cannot simplify the collision operator and
the solution cannot be given as explicitly as in Region I, but it can
be found numerically. Here, we only need to use that, due to the
results of Section \ref{sec:AverageMagDrift}, equation
\eq{eq:DKEionsOrbitAveraged} gives
\begin{eqnarray}\label{eq:SizeGRegionII}
 G_i^{\mathrm{II}} \sim |\alpha|^{1/2}\nu_{*i}^{-1}\epsilon_i F_{i0},
\end{eqnarray}
\begin{eqnarray}
G_i^{\mathrm{III}} \sim |\alpha|^{1/2} \nu_{*i}^{-1}\epsilon_i F_{i0}
\end{eqnarray}
and also
\begin{eqnarray}
\partial_\lambda G_i^{\mathrm{II}} \sim |\alpha|^{1/2} \nu_{*i}^{-1}\epsilon_i F_{i0},
\end{eqnarray}
\begin{eqnarray}
  \partial_\lambda G_i^{\mathrm{III}} \sim |\alpha|^{1/2} \nu_{*i}^{-1}\epsilon_i F_{i0}.
\end{eqnarray}
Now, recall that at the end of Section \ref{sec:breakdownOfExpansion}
we proved that $G_i^{[-1]} = 0$ for passing particles to lowest order
in $\alpha$. Therefore, the boundary condition needed to solve for
$G_i^{\mathrm{II}}$ is $G_i^{\mathrm{II}}=0$ at the passing/trapped
interface. Finally, $G_i^{\mathrm{III}}$ is determined by imposing
continuity between Regions II and III (we cannot
  discard the existence of a discontinuity in $\partial_\lambda G_i$
  between Regions II and III).

The relation between the different regions is provided by the
emergence of a collisional layer in Region IV, the thin gray area in
figure \ref{fig:EnergyContours}. In this region the bounce-averaged
equation \eq{eq:DKEionsOrbitAveraged} is not suitable because
particles collide too frequently. In the layer the parallel streaming
and collision terms in the drift-kinetic equation
\eq{eq:FPG1dimensionful} balance each other, giving
\begin{equation}
v_{||}\bun\cdot\nabla \sim \nu_i\partial_\xi^2.
\end{equation}
Equivalently,
\begin{equation}\label{eq:sizeCollisionalLayer}
v_{||}\bun\cdot\nabla \sim \frac{\nu_i}{\delta \xi^2},
\end{equation}
where $\xi = v_{||}/v_{ti}$ and $\delta \xi$ stands for the width of
the layer. The distribution function has large parallel velocity
derivatives in the collisional layer, and the pitch-angle scattering
piece of the collision operator dominates. The secondary well has a
typical size $L_1 \sim |\alpha| L_0$ and particles trapped in it have
typical parallel velocities $v_{||}\sim |\alpha|^{1/2} v_{t i}$. Then,
\eq{eq:sizeCollisionalLayer} yields
\begin{equation}\label{eq:sizeCollisionalLayerLeft}
  \delta\xi\sim
  \left(\frac{\nu_i L_1}{|\alpha|^{1/2}v_{ti}}\right)^{1/2}\sim
  \left(\frac{\nu_i L_1}{|\alpha|^{3/2}v_{ti}}\right)^{1/2}
|\alpha|^{1/2}\ll |\alpha|^{1/2},
\end{equation}
where we have used \eq{eq:LowCollCondSmallWell}. Hence, the width in
$v_{||}$ of the collisional layer around the left lobe of Region IV
(see figure \ref{fig:EnergyContours}) is much smaller than the typical
value of $v_{||}$ at the boundary, as it should. As for the right lobe
of Region IV, with size $L_0$ and typical parallel velocities
$v_{||}\sim v_{ti}$, one also gets that the width of the layer is
consistent,
\begin{equation}\label{eq:sizeCollisionalLayerRight}
  \delta\xi\sim
  \left(\frac{\nu_i L_0}{v_{ti}}\right)^{1/2} \ll 1,
\end{equation}
due to \eq{eq:LowCollCond}.

The equation in the collisional boundary layer is
\begin{equation}
  v_{||}\bun\cdot\nabla G_i^{\mathrm{IV}} = 
\frac{v_{||}}{2\varepsilon \tau_\perp}\frac{{\cal B}_0}{B}\partial_\lambda
\left(v_{||}\lambda\partial_\lambda G_i^{\mathrm{IV}}\right).
\end{equation}
Multiplying by $v_{||}^{-1}$ and integrating along the field line,
\begin{equation}
\oint
\frac{ {\cal B}_0}{2\varepsilon \tau_\perp\bB\cdot\nabla\Theta}
\partial_\lambda
\left(v_{||}\lambda\partial_\lambda G_i^{\mathrm{IV}}\right)\dd \Theta
=0,
\end{equation}
where $\oint$ stands for the integral over the corresponding trapped
orbit.  Hence, integrating in $\lambda$ over the collisional layer, we
get
\begin{eqnarray}\label{eq:MatchingCondition}
\fl
\oint
\frac{{\cal B}_0}{2\varepsilon \tau_\perp \bB\cdot\nabla\Theta}
\left(v_{||}\lambda\partial_\lambda G_i^{I}\right)\dd\Theta
+
\oint
\frac{{\cal B}_0}{2\varepsilon \tau_\perp \bB\cdot\nabla\Theta}
\left(v_{||}\lambda\partial_\lambda G_i^{\mathrm{II}}\right)\dd\Theta
\nonumber\\[5pt]
\fl\hspace{1cm}
+
\oint
\frac{{\cal B}_0}{2\varepsilon \tau_\perp \bB\cdot\nabla\Theta}
\left(v_{||}\lambda\partial_\lambda G_i^{\mathrm{III}}\right)
\dd\Theta
=
0,
\end{eqnarray}
where all terms are evaluated at the boundary of Region IV.

We already know that $\partial_\lambda G_i^{\mathrm{I}} \sim
  \nu_{*i}^{-1}\epsilon_i F_{i0}$. Taking into account that
  in Region I $v_{||}\sim |\alpha|^{1/2}v_{ti}$, and the size of the
  secondary well is $L_1\sim|\alpha| L_0$, we deduce that the first
  term in \eq{eq:MatchingCondition} is $O(|\alpha|^{3/2})$. Therefore,
\begin{eqnarray}\label{eq:MatchingCondition2}
\fl
\oint
\frac{{\cal B}_0}{2\varepsilon \tau_\perp \bB\cdot\nabla\Theta}
\left(v_{||}\lambda\partial_\lambda G_i^{\mathrm{II}}\right)\dd\Theta
\nonumber\\[5pt]
\fl\hspace{1cm}
+
\oint
\frac{{\cal B}_0}{2\varepsilon \tau_\perp \bB\cdot\nabla\Theta}
\left(v_{||}\lambda\partial_\lambda G_i^{\mathrm{III}}\right)\dd\Theta
=
O(|\alpha|^{3/2}).
\end{eqnarray}
Since the typical value of the parallel velocity in the terms on the
left side of the previous equation is $v_{||}\sim v_{ti}$, we infer
that the jump of $\partial_\lambda G_i$ between Regions II and III is
$O(|\alpha|^{3/2}\nu_{*i}^{-1}\epsilon_i F_{i
    0})$. It may seem that this jump is negligible compared to
$\partial_\lambda G_i\sim |\alpha|^{1/2}\nu_{*i}^{-1}\epsilon_i
F_{i0}$, but in general there is a number of small wells of order
$|\alpha|^{-1}$ in a field line and the accumulation of these
discontinuities modifies $G_i$ by a quantity of order
$|\alpha|^{1/2}\nu_{*i}^{-1}\epsilon_i F_{i 0}$.

We denote the union of Regions II and III by Region
  II$'$. The smallness of the size of the boundary layer (Region IV)
  implies that $G_i^{\mathrm{I}}$ and $G_i^{\mathrm{II}'}$ are
  continuous to lowest order (their derivatives are not). Since we
  know $\partial_\lambda G_i^{\mathrm{I}}$, we can integrate from the
  boundary of the secondary well towards the bottom $\lambda_b$. It is
  easy to realize that the change in $G_i^{\mathrm{I}}$ is negligible
  compared to $G_i^{\mathrm{II}'}$, giving $G_i^{\mathrm{I}}\sim
  G_i^{\mathrm{II}'}\sim |\alpha|^{1/2}\nu_{*i}^{-1}\epsilon_i F_{i
    0}$. All these scalings work analogously for electrons.

Using the above results and noting that in velocity
  space the fraction of particles trapped in secondary wells scales
as $|\alpha|^{1/2}$, we find that all trapped trajectories contribute
with the same scaling to the flux-surface averaged radial electric
current,
\begin{eqnarray}\label{eq:linearScalingAgain}
\fl
\left\langle
    \bJ\cdot\nabla\psi
\right\rangle_\psi 
=
\left\langle
\int G_i^{[-1]}\overline{v_{\psi,\spe}^{(0)}}\dd^3v
\right\rangle_\psi + \dots
=
\nonumber\\[5pt]
\fl\hspace{1cm}
\left\langle
\int G_i^{\mathrm{I}}\overline{v_{\psi,\spe}^{(0)}}\dd^3v
\right\rangle_\psi
+
\left\langle
\int G_i^{\mathrm{II}'}\overline{v_{\psi,\spe}^{(0)}}\dd^3v
\right\rangle_\psi
 + \dots
\sim
\frac{|\alpha| \epsilon_i^2 v_{ti}}{\nu_i L_0} e n_i v_{ti} |\nabla\psi|,
\end{eqnarray}
where the dots stand for higher-order terms. Here, we have also
employed that the size of the well in $\chi$ is of order $|\alpha|$,
that the number of small wells on a given magnetic field line is of
order $|\alpha|^{-1}$, and that the number of lines with small wells
is also of order $|\alpha|^{-1}$.

The main result obtained in this paper, equation
\eq{eq:linearScalingAgain}, contradicts the $|\alpha|^{3/2}$ scaling
typically associated to a symmetric magnetic field where a ripple
magnetic field that creates secondary wells has been
added~\cite{Ho1987}. The argument leading to such a scaling is easy
and plausible, but the above rigorous treatment shows that it is
incorrect. The former is based on the assumption that the secondary
wells dominate transport, whereas we have proven that all trapped
trajectories contribute the same. Going through similar steps as
above, it is easy to show that the scaling $|\alpha|^{3/2}$ can be
obtained by adopting the assumption that particles trapped in the
large wells of $\bB_0$ drift outwards at an unrealistically low rate
$O(\alpha\epsilon_i v_{ti})$.

Finally, one might wonder how the rotation criterion
  \eq{eq:rotationcriterion} is modified if large helicity
  perturbations are present. When $\alpha\bun_0\cdot\nabla B_1
  \sim \bun_0\cdot\nabla B_0$, we have only treated the $1/\nu$
  regime, and therefore the comparison has to be carried out with
  \eq{eq:rotationcriterion} particularized for low collisionality. If
  one uses \eq{eq:klowcollisionality}, condition
  \eq{eq:rotationcriterion} can be more precisely formulated, giving
  the rotation criterion
\begin{equation}
|\alpha| < \sqrt{\nu_{*i}\epsilon_i}
\end{equation}
for small helicity perturbations. The same arguments that lead to this
criterion, exposed in reference \cite{CalvoParraVelascoAlonso13}, can
be repeated employing now the scaling \eq{eq:linearScalingAgain},
yielding the rotation criterion
\begin{equation}
|\alpha| < \nu_{*i}\epsilon_i
\end{equation}
for large helicity perturbations. Then, for a fixed value of $\alpha$,
flow damping is stronger if large helicity perturbations are not
avoided. This should be taken into account in future quasisymmetric
stellarator designs.

\section{Conclusions}
\label{eq:Conclusions}

Quasisymmetry is an interesting design concept in stellarator research
but it is known~\cite{Garren1991} that it cannot be achieved exactly,
even if configurations reasonably close to quasisymmetric are
feasible~\cite{Anderson1995}. In the light of these facts, it is
important to understand quantitatively how physical
features associated to quasisymmetry, such as the possibility to have
large equilibrium flows, are affected by small deviations from it. To
answer this question, we have calculated how the flux-surface averaged
radial electric current (that identically vanishes for a
quasisymmetric configuration) deviates from zero depending on the size
and other properties of the non-quasisymmetric magnetic field
perturbation.

The systematic treatment of the problem started in reference
\cite{CalvoParraVelascoAlonso13}, where the scaling with the size of
the perturbation was derived for the least deleterious type of
perturbation, one with small spatial gradients. Let the magnetic field
be $\bB = \bB_0 + \alpha \bB_1$, where $\bB_0$ is quasisymmetric and
$\alpha \bB_1$ a perturbation. If the helicity of the latter is
sufficiently small (see the Introduction for a precise statement of
the conditions), then
\begin{equation}\label{eq:quadraticScalingConclusions}
  \left\langle
    \bJ\cdot\nabla\psi
  \right\rangle_\psi\sim\alpha^2 k,
\end{equation}
where the form of the factor $k$ depends on the collisionality regime.

More dangerous types of perturbations (again, we refer the reader to
the Introduction for the technical details), that should be avoided,
if possible, when designing quasisymmetric stellarators, have been the
subject of this paper. In general, when the gradient of the
perturbation is large one has
\begin{equation}\label{eq:WorstScalingConclusions}
  \left\langle
    \bJ\cdot\nabla\psi
  \right\rangle_\psi\sim O(\alpha^0),
\end{equation}
which amounts to say that the quasisymmetric properties of $\bB_0$
have been lost. However, an intermediate situation between
\eq{eq:quadraticScalingConclusions} and
\eq{eq:WorstScalingConclusions} exists when the gradient of the
perturbation is large but it is aligned with the magnetic field
lines. Then, one gets
\begin{equation}\label{eq:LinearScalingConclusions}
  \left\langle
    \bJ\cdot\nabla\psi
  \right\rangle_\psi\sim \frac{|\alpha| \epsilon_i^2 v_{ti}}{\nu_i L_0}
 e n_i v_{ti}|\nabla\psi|
\end{equation}
in the $1/\nu$ regime. Such large parallel derivatives are typically
associated to the appearance of small secondary wells and it has often
been believed in the literature that these small wells dominate
transport and that they produce a scaling $|\alpha|^{3/2}$. We have
shown that the $|\alpha|^{3/2}$ scaling is incorrect.

We have also explained why \eq{eq:LinearScalingConclusions} implies
that the capability of the stellarator to rotate is reduced with
respect to the case in which only small helicity perturbations exist.

\ack

This work was supported by EURATOM and carried out within the
framework of the EUROfusion Consortium. This project has received
funding from the EU Horizon 2020 research and innovation
programme. The views and opinions expressed herein do not necessarily
reflect those of the European Commission. This research was supported
in part by grant ENE2012-30832, Ministerio de Econom\'{\i}a y
Competitividad, Spain.

\section*{References}

\end{document}